\documentclass[11pt]{article}
\usepackage{times}
\usepackage[pagebackref]{hyperref}
\usepackage{graphicx,amsthm,amsmath,amsfonts,amssymb, latexsym}
\usepackage[usenames,dvipsnames]{color}
\usepackage[margin=1in]{geometry}

\title{{\bf Combinatorial limitations of average-radius list-decoding}}

\author{{\large {\rm Venkatesan Guruswami}}\footnote{Research supported in
    part by NSF grants CCF 0953155 and CCF 0963975. Email: {\tt 
    guruswami@cmu.edu},  {\tt srivatsa@cs.cmu.edu}} 
    \and {\rm {\large Srivatsan Narayanan}}\footnotemark[\value{footnote}]}
    \date{Computer Science Department \\
    Carnegie Mellon University \\
    Pittsburgh, PA 15213. ~ \\~\\
    April 2013}

\newtheorem{theorem}{Theorem}
\newtheorem{fact}[theorem]{Fact}
\newtheorem{lemma}[theorem]{Lemma}
\newtheorem{proposition}[theorem]{Proposition}
\newtheorem{definition}[theorem]{Definition}
\newenvironment{remark}{\noindent {{\em Remark}.}}{$\Box$ \medskip}
\renewenvironment{proof}{\noindent {\bf Proof:}}{\hfill$\Box$}
\newenvironment{proofof}[1]{\noindent {\bf Proof of {#1}:}}{\hfill$\Box$}

\renewcommand{\le}{\leqslant}
\renewcommand{\leq}{\leqslant}
\renewcommand{\ge}{\geqslant}
\renewcommand{\geq}{\geqslant}

\renewcommand{\P}{{\bf Pr}}
\renewcommand{\Pr}{{\bf Pr}}
\newcommand{\I}{\mathbb{I}}
\newcommand{\E}{{\bf E}}
\newcommand{\Var}{{\bf Var}}

\newcommand{\eps}{\varepsilon}
\newcommand{\h}{h}
\newcommand{\Reals}{\mathbb{R}}
\newcommand{\F}{{\bf F}}

\newcommand{\wt}{\operatorname{wt}}
\newcommand{\vecspan}{\operatorname{Span}}
\newcommand{\supp}{\operatorname{Supp}}
\newcommand{\suppstar}{\supp^{\ast}}
\newcommand{\ball}{{\bf B}}
\newcommand{\zerovec}{{\bf 0}}
\newcommand{\dist}{d}

\newcommand{\zo}{\{ 0, 1 \}}
\newcommand{\zq}{[q]}
\newcommand{\erasym}{?}

\newcommand{\sfavg}{\text{\sffamily{avg}}}
\newcommand{\setcomplement}[1]{\overline{#1}}

\newcommand{\List}{\mathcal{L}}
\newcommand{\Dmax}{D_{\text{\sffamily{max}}}}
\newcommand{\Davg}{D_{\text{\sffamily{avg}}}}

\begin{document}

\maketitle
\thispagestyle{empty}

\begin{abstract}
We study certain combinatorial aspects of list-decoding, motivated by the
exponential gap between the known upper bound (of $O(1/\gamma)$) and lower
bound (of $\Omega_p (\log (1/\gamma))$) for the list-size needed to list
decode up to error fraction $p$ with rate $\gamma$ away from capacity, i.e.,
$1-\h(p)-\gamma$ (here $p\in (0, \frac{1}{2})$ and $\gamma > 0$). Our main
result is the following:

\begin{itemize}

\item We prove that in any binary code $C \subseteq \zo^n$ of rate
$1- \h(p) - \gamma$, there must exist a set $\List \subset C$ of $\Omega_p
(1/\sqrt{\gamma})$ codewords such that the average distance of the points in
$\List$ from their centroid is at most $pn$. In other words, there must exist
$\Omega_p(1/\sqrt{\gamma})$ codewords with low ``average radius.''

\smallskip
The standard notion of list-decoding corresponds to working with the {\em
maximum} distance of a collection of codewords from a center instead of {\em
average} distance.  The average-radius form is in itself quite natural; for
instance, the classical Johnson bound in fact implies average-radius
list-decodability.
\end{itemize}

\noindent
The remaining results concern the standard notion of list-decoding, and help
clarify the current state of affairs regarding combinatorial bounds for
list-decoding:

\begin{itemize}
\item We give a short simple proof, over all fixed alphabets, of the
    above-mentioned $\Omega_p(\log (1/\gamma))$ lower bound. Earlier, this
    bound followed from a complicated, more general result of Blinovsky.

\item We show that one {\em cannot} improve the $\Omega_p(\log (1/\gamma))$
    lower bound via techniques based on identifying the zero-rate regime for
    list-decoding of constant-weight codes (this is a typical approach for
    negative results in coding theory, including the $\Omega_p
    (\log (1/\gamma))$ list-size lower bound). On a positive note, our
    $\Omega_p(1/\sqrt{\gamma})$ lower bound for average-radius list-decoding
    circumvents this barrier.

\item We exhibit a ``reverse connection'' between the existence of
    constant-weight and general codes for list-decoding, showing that the best
    possible list-size, as a function of the gap $\gamma$ of the rate to the
    capacity limit, is the same up to constant factors for both
    constant-weight codes (with weight bounded away from $p$) 
    and general codes.

\item We give simple second moment based proofs that w.h.p.\ a list-size of
    $\Omega_p (1/\gamma)$ is needed for list-decoding {\em random} codes from
    errors as well as erasures. For {\em random linear} codes, the
    corresponding list-size bounds are
    $\Omega_p (1/\gamma)$ for errors and $\exp(\Omega_{p}(1/\gamma))$ for
    erasures.
\end{itemize}
\end{abstract}

\newpage

% =========================

\section{Introduction}

The list-decoding problem for an error-correcting code $C \subseteq \Sigma^n$
consists of finding the set of all codewords of $C$ with Hamming distance at
most $pn$ from an input string $y \in \Sigma^n$. Though it was originally
introduced in early work of Elias and Wozencraft~\cite{elias,wozencraft} in
the context of estimating the decoding error probability for random error
models, recently the main interest in list-decoding has been for adversarial
error models. List decoding enables correcting up to a factor two more
worst-case errors compared to algorithms that are always restricted to output
a unique answer, and this potential has even been realized
algorithmically~\cite{GR-FRS,gur-frs-lin-alg}.

In this work, we are interested in some fundamental combinatorial questions
concerning list-decoding, which highlight the important tradeoffs in this
model. Fix $p \in (0, \frac{1}{2})$ and a positive integer $L$. We say that a
binary code $C \subseteq \zo^n$ is $(p, L)$ list-decodable if every Hamming
ball of radius $pn$ has {\em less than} $L$ codewords. Here, $p$ corresponds
to the error-fraction and $L$ to the list-size needed by the error-correction
algorithm. Note that $(p, L)$ list-decodability imposes a sparsity requirement
on the distribution of codewords in the Hamming space. A natural combinatorial
question that arises in this context is to place bounds on the largest size of
a code meeting this requirement. In particular, an outstanding open question
is to characterize the maximum rate (defined to be the limiting ratio
$\frac{1}{n} \log |C|$ as $n \to \infty$) of a $(p, L)$ list-decodable code.

By a simple volume packing argument, it can be shown that a $(p, L)$
list-decodable code has rate at most $1-\h(p)+o(1)$. (Throughout, for $z \in
[0,\frac{1}{2}]$, we use $\h(z)$ to denote the binary entropy function at
$z$.) Indeed, picking a random center $x$, the Hamming ball $\ball(x,pn)$
contains at least $|C| \cdot \binom{n}{pn} 2^{-n}$ codewords in expectation.
Bounding this by $(L-1)$, we get the claim. On the positive side, in the limit
of large $L$, the rate of a $(p, L)$ list-decodable code approaches the
optimal $1-\h(p)$. More precisely, for any $\gamma > 0$, there exists a
$(p, 1/\gamma)$ list-decodable code of rate at least $1- \h(p) -\gamma$. In
fact, a random code of rate $1-\h(p)-\gamma$ is $(p, 1/\gamma)$ list-decodable
w.h.p.~\cite{ZP81,elias91}, and a similar result holds for random linear codes
(with list-size $O_p (1/\gamma)$)~\cite{GHK11}. In other words, a dense random
packing of $2^{(1-\h(p)-\gamma)n}$ Hamming balls of radius $pn$ (and therefore
volume $\approx 2^{\h(p)n}$ each) is ``near-perfect'' w.h.p.\ in the sense
that no point is covered by more than $O_p(1/\gamma)$ balls.

The determination of the best asymptotic code rate of binary $(p, L)$
list-decodable codes as $p,L$ are held fixed and the block length grows is
wide open for every choice of $p \in (0, \frac{1}{2})$ and integer $L \ge 1$.
However, we {\em do} know that for each fixed $p \in (0, \frac{1}{2})$, this
rate approaches $1- \h(p)$ in the limit as $L \to \infty$. To understand this
rate of convergence as a function of list-size $L$, following~\cite{GHK11},
let us define $L_{p, \gamma}$ to be the minimum integer $L$ such that there
exist $(p, L)$ list-decodable codes of rate $1-\h(p)-\gamma$ for infinitely
many block lengths $n$ (the quantity $\gamma$ is the ``gap'' to
``list-decoding capacity''). In \cite{blinovsky}, Blinovsky showed that a
$(p, L)$ list-decodable code has rate at most $1-\h(p)-2^{-\Theta_p(L)}$. In
particular, this implies that for any finite $L$, a $(p, L)$ list-decodable
code has rate strictly below the optimal $1-\h(p)$. Stated in terms of
$L_{p, \gamma}$, his result implies the corollary
$L_{p, \gamma} \geq \Omega_p(\log (1/\gamma))$ for rates $\gamma$-close to
capacity.  We provide a short and simple proof of this corollary in
Section~\ref{sec::std-ld}. Our proof works almost as easily over non-binary
alphabets. (Blinovsky's subsequent proof for the non-binary case
in~\cite{blin-q-ary,blin-convexity} involved substantial technical effort.
However, his results also give non-trivial bounds for every finite $L$, as
opposed to just the growth rate of $L_{p, \gamma}$.)

Observe the exponential gap (in terms of the dependence on $\gamma$) between
the $O(1/\gamma)$ upper bound and $\Omega_p(\log(1/\gamma))$ lower bounds on
the quantity $L_{p, \gamma}$. Despite being a basic and fundamental question
about sphere packings in the Hamming space and its direct relevance to
list-decoding, there has been no progress on narrowing this asymptotic gap in
the 25 years since the works of Zyablov-Pinsker~\cite{ZP81} and
Blinovsky~\cite{blinovsky}. This is the motivating challenge driving this
work.

% ===============

\subsection{Prior work on list-size lower bounds}

We now discuss some lower bounds (besides Blinovsky's general lower bound) on
list-size that have been obtained in restricted cases.

Rudra shows that the $O_p(1/\gamma)$ bound obtained via the probabilistic
method for random codes is, in fact, tight up to constant
factors~\cite{atri-ieeeit}. Formally, there exists $L = \Omega_p(1/\gamma)$
such that a random code of rate $1-\h(p)-\gamma$ is {\em not} $(p, L)$
list-decodable w.h.p. His proof uses near-capacity-achieving codes for the
binary symmetric channel, the existence of which is promised by Shannon's
theorem, followed by a second moment argument.  We give a simpler proof of
this result via a more direct use of the second moment method. This has the
advantage that it works uniformly for random general as well as random linear
codes, and for channels that introduce errors as well as erasures.

Guruswami and Vadhan~\cite{GV} consider the problem of list-size bounds when
the channel may corrupt close to half the bits, that is, when
$p = \frac{1}{2} -\eps$, and more generally $p= 1- 1/q- \eps$ for codes over
an alphabet of size $q$. (Note that decoding is impossible if the channel
could corrupt up to a half fraction of bits.) They show that there exists
$c > 0$ such that for all $\eps > 0$ and all block lengths $n$, any
$(\frac{1}{2} - \eps, c/\eps^2)$ list-decodable code contains $O_\eps(1)$
codewords. For $p$ bounded away from $\frac{1}{2}$ (or $1-1/q$ in the
$q$-ary case), their methods do not yield any nontrivial list-size lower
bound as a function of gap $\gamma$ to list-decoding capacity.

% ===============

\subsection{Our main results}

We have already mentioned our new proof of the $\Omega(\log (1/\gamma))$
list-size lower bound for list-decoding general codes, and the asymptotically
optimal list-size lower bound for random (and random linear) codes.

Our main result concerns an average-radius variant of list-decoding. This
variant was implicitly used in \cite{blinovsky,GV} en route their list-size
lower bounds for standard list-decoding. In this work, we formally abstract
this notion: a code is {\em $(p, L)$ average-radius list-decodable} if for
every $L$ codewords, the {\em average} distance of their centroid from the $L$
codewords exceeds $pn$. Note that this is a stronger requirement than $(p, L)$
list-decodability where only the {\em maximum} distance from any center point
to the $L$ codewords must exceed $pn$.

We are able to prove nearly tight bounds on the achievable rate of a $(p, L)$
average-radius list-decodable code. To state our result formally, denote by
$L^{\sfavg}_{p,\gamma}$ the minimum $L$ such that there exists a $(p, L)$
average-radius list-decodable code family of rate $1-\h(p)-\gamma$. A simple
random coding argument shows that a random code of $1-\h(p)-\gamma$ is
$(p, 1/\gamma)$ average-radius list-decodable (matching the list-decodability
of random codes). That is, $L^{\sfavg}_{p,\gamma} \leq 1/\gamma$. Our main
technical result is a lower bound on the list-size that is polynomially
related to the upper bound, namely $L^{\sfavg}_{p, \gamma} \geq
\Omega_p (\gamma^{-1/2})$.

We remark that the classical Johnson bound in coding theory in fact proves
the average-radius list-decodability of codes with good minimum distance
---namely, a binary code of relative distance $\delta$ is
$(J(\delta - \delta/L), L)$ average-radius list-decodable, where
$J(z) = (1-\sqrt{1-2z})/2$ for $z \in [0,\frac{1}{2}]$. (This follows from a
direct inspection of the proof of the Johnson bound~\cite{GS-johnson}.) Also,
one can show that if a binary code is $(\frac{1}{2}- 2^i \eps,
O(1/(2^{2i}\eps^2))$ list-decodable for all $i=0, 1, 2, \ldots$, then it is
also $(\frac{1}{2}-2\eps, O(1/\eps^2))$ average-radius
list-decodable~\cite{CG-rip-ld}. This shows that at least in the high noise
regime, there is some reduction between these notions.  Further, a suitable
soft version of average-radius list-decodability can be used to construct
matrices with a certain restricted isometry property~\cite{CG-rip-ld}. For
these reasons, we feel that average-radius list-decodability is a natural
notion to study, even beyond treating it as a vehicle to understand (standard)
list-decoding. In fact, somewhat surprisingly, one of our {\em constructions}
of traditional list-decodable codes with a strong weight requirement on the
codewords proceeds naturally via average-radius list-decodability; see
Theorem~\ref{thm::zrate} and the discussion following it for details.

% ===============

\subsection{Our other results} \label{subsec::otherresults}

We also prove several other results that clarify the landscape of
combinatorial limitations of list-decodable codes. Many results showing rate
limitations in coding theory proceed via a typical approach in which they pass
to a constant weight $\lambda \in (p, \frac{1}{2}]$; i.e., they restrict the
codewords to be of weight exactly $\lambda n$. They show that under this
restriction, a code with the stated properties must have a constant number of
codewords (that is, asymptotically {\em zero rate}). Mapping this bound back
to the unrestricted setting one gets a rate upper bound of $1 - \h(\lambda)$
for the original problem. For instance, the Elias-Bassalygo bound for rate
$R$ vs.\ relative distance $\delta$ is of this nature (here $\lambda$ is
picked to be the Johnson radius for list-decoding for codes of relative
distance $\delta$).

The above is also the approach taken in Blinovsky's work~\cite{blinovsky} as
well as that of \cite{GV}. We show that such an approach does not and
{\em cannot} give any bound better than Blinovsky's
$\Omega_p (\log (1/\gamma))$ bound for $L_{p,\gamma}$. More precisely, for
any $\lambda \geq p + 2^{- b_p L}$ for some $b_p > 0$, we show that there
exists a $(p, L)$ (average-radius) list-decodable code of rate
$\Omega_{p, L}(1)$. Thus in order to improve the lower bound, we {\em must}
be able to handle codes of strictly positive rate, and cannot deduce the bound
by pinning down the zero-rate regime of constant-weight codes. This perhaps
points to why improvements to Blinovsky's bounds have been difficult. On a
positive note, we remark that we {\em are} able to effect such a proof for
average-radius list-decoding (some details follow next).

To describe the method underlying our list-size lower bound for average-radius
list-decoding, it is convenient to express the statement as an upper bound on
rate in terms of list-size $L$. Note that a list-size lower bound of
$L \ge \Omega_p(1/\sqrt{\gamma})$ for $(p, L)$ average-radius list-decodable
codes of rate $1-\h(p) - \gamma$ amounts to proving an upper bound of
$1-\h(p)- \Omega_p(1/L^2)$ on the rate of $(p, L)$ average-radius
list-decodable codes. Our proof of such an upper bound proceeds by first
showing a rate upper bound of $\h(\lambda)-\h(p)- a_p/L^2$ for such codes when
the codewords are all restricted to all have weight $\lambda n$, for a
suitable choice of $\lambda$, namely $\lambda = p + a'_p/L$. To map this bound
back to the original setting (with no weight restrictions on codewords), one
simply notes that every $(p, L)$ average-radius list-decodable code of rate
$R$ contains as a subcode, a translate of a constant $\lambda n$-weight
code of rate $R- (1-\h(\lambda))$. (The second step uses a well-known
argument.)

Generally speaking, by passing to a constant-weight subcode, one can translate
combinatorial results on limitations of constant-weight codes to results
showing limitations for the case of general codes. But this leaves open the
possibility that the problem of showing limitations of constant-weight codes
may be harder than the corresponding problem for general codes, or worse
still, have a different answer making it impossible to solve the problem for
general codes via the methodology of passing to constant-weight codes. We show
that for the problem of list-decoding this is fortunately not the case, and
there is, in fact, a ``reverse connection'' of the following form: A rate
upper bound of $1-\h(p)-\gamma$ for $(p, L)$ list-decodable codes implies a
rate upper bound of
$\h(\lambda)- \h(p) - \left( \frac{\lambda-p}{\frac{1}{2}-p}\right) \gamma$
for $(p, L)$ list-decodable codes whose codewords must all have Hamming
weight $\lambda n$. A similar claim holds also for average-radius
list-decodability, though we don't state it formally.

% ===============

\subsection{Our proof techniques}

Our proofs in this paper employ variants of the standard probabilistic method.
We show an extremely simple probabilistic argument that yields a
$\Omega_p (\log (1/\gamma))$ bound on the list-size of a standard
list-decodable code; we emphasize that this is qualitatively the tightest
known bound in this regime.

For the ``average-radius list-decoding'' problem that we introduce, we are
able to improve this list-size bound to $\Omega_p (1/\sqrt{\gamma})$. The
proof is based on the idea that instead of picking the ``bad list-decoding
center'' $x$ uniformly at random, one can try to pick it randomly very close
to a designated codeword $c^\ast$, and this still gives similar guarantees on
the number of near-by codewords. Now since the quantity of interest is the
average radius, this close-by codeword gives enough savings for us. In order
to estimate the probability that a typical codeword $c$ belongs to the list
around $x$, we write this probability explicitly as a function of the Hamming
distance between $c^\ast$ and $c$, which is then lower bounded using
properties of hypergeometric distributions and Taylor approximations for the
binary entropy function.

For limitations of list-decoding random codes, we define a random variable
$W$ that counts the number of ``violations'' of the list-decoding property of
the code. We then show that $W$ has a exponentially large mean, around which
it is concentrated w.h.p. This yields that the code cannot be list-decodable
with high probability, for suitable values of rate and list-size parameters.

% ===============

\subsection{Organization}

We define some useful notation and the formal notion of average-radius
list-decodability in Section~\ref{sec::prelim}. Our main list-size lower bound
for average-radius list-decoding appears in Section~\ref{sec::strongld}. We
give our short proof of Blinovsky's lower bounds for binary and general
alphabets in Section~\ref{sec::std-ld}.  Our results about the zero-error rate
regime for constant-weight codes and the reverse connection between
list-decoding bounds for general codes and constant-weight codes appear in
Section~\ref{sec::constwt}. Finally, our list-size lower bounds for random
codes are stated in Section~\ref{sec::random}; for reasons of space, the
proofs for these bounds appear in the appendix.

% =========================

\section{Preliminaries and notation} \label{sec::prelim}
\subsection{List decoding}

We recall some standard terminology regarding error-correcting codes.

Let $[n]$ denote the index set $\{ 1, 2, \ldots, n \}$. For $q \geq 2$, let
$\zq$ denote the set $\{ 0, 1, \ldots, q-1 \}$. A {\em $q$-ary code} refers to
any subset $C \subseteq \zq^n$, where $n$ is the {\em blocklength} of $C$. We
will mainly focus on the special case of binary codes corresponding to $q =
2$. The rate $R = R(C)$ is defined to be $\frac {\log |C|}{n \log q}$. For
$x \in \zq^n$ and $S \subseteq [n]$, the restriction of $x$ to the coordinates
in $S$ is denoted $x|_S$. Let $\supp (x) := \{ i \in [n] ~:~ x_i \neq 0 \}$.
A {\em subcode} of $C$ is a subset $C'$ of $C$. We say that $C$ is a {\em
constant-weight code} with weight $w \in [0, n]$, if all its codewords have
weight exactly $w$. (Such codes are studied in Section~\ref{sec::constwt}.)

For $x, y \in \zq^n$, define the {\em Hamming distance} between $x$ and $y$,
denoted $\dist (x, y)$, to be the number of coordinates in which $x$ and $y$
differ. The {\em (Hamming) weight} of $x$, denoted $\wt (x)$, is $\dist
(\zerovec, x)$, where $\zerovec$ is the vector in $\zq^n$ with zeroes in all
coordinates. The {\em (Hamming) ball} of radius $r$ centered at $x$, denoted
$\ball (x,r)$, is the set $\{ y \in \zq^n ~:~ \dist (x, y) \leq r \}$. In this
paper, we also need the following notions of distance of a (small)
``list'' $\List$ of vectors from a ``center'' $x$:
\begin{definition}
Given a center $x \in \zq^n$ and a nonempty list $\List \subseteq \zq^n$,
define the maximum and average distances of $\List$ from $x$ respectively by:
\begin{align*}
\Dmax (x, \List) &:= \max \{ \dist (x, c) ~:~ c \in \List \} ,\text{ and} \\
\Davg(x, \List) &:= \E_{c \in \List} \ \Big[ \dist (x, c) \Big] =
    \frac{1}{|\List|} \sum_{c \in \List} \dist (x, c)
.\end{align*}
\end{definition}
It is well-known (cf., e.g.,~Lemma~5 in~\cite{GV}) that the average-radius of
a list is minimized by the {\em coordinatewise majority} (or {\em centroid})
of the list:
\begin{fact} \label{fact::centroid}
Let $\List = \{ c_1, c_2, \ldots, c_L \} \subseteq \zo^n$ be an arbitrary list
of codewords, and let $a \in \zo^n$ be its {\em centroid}; that is, for any
coordinate $j$, the $j^{\rm th}$ entry of $a$ is the majority of the
corresponding entries of $c_1, c_2, \ldots, c_L$ (breaking ties arbitrarily).
Then
\[
\Davg (a, \List) = \min_{a' \in \zo^n} \Davg (a', \List)
.\]
\end{fact}

Next, we formalize the error recovery capability of the code using
list-decoding.
\begin{definition} \label{def::ld}
Fix $0 < p < \frac{1}{2}$ and a positive integer $L$. Let $C$ be a $q$-ary
code with blocklength $n$.
\begin{enumerate}
\itemsep=0ex

\item \label{itm::stdld} $C$ is said to be {\em $(p, L)$ list-decodable} if
    for all $x \in\zq^n$, the ball $\ball(x, pn)$ contains at most $L-1$
    codewords of $C$. Equivalently, for any $x$ and any list $\List \subseteq
    C$ of size at least $L$, we have $\Dmax(x, \List) > pn$.

\item $C$ is said to be {\em $(p, L)$ average-radius list-decodable} if for
    any center $x$ and any $L$-tuple $\List$ of codewords, we have $\Davg(x,
    \List) > pn$.
\end{enumerate}
\end{definition}

For constant-weight codes, it is convenient to augment the above notation with
the weight parameter:
\begin{definition} \label{def::ld-constweight}
Let $p, L, q, n, C$ be as in Definition~\ref{def::ld}, and let $0 < \lambda
\leq \frac{1}{2}$. $C$ is said to be {\em $(\lambda; p, L)$ (average-radius)
list-decodable} if $C$ is  $(p, L)$ (average-radius) list-decodable, and every
codeword in $C$ has weight exactly $\lambda n$.
\end{definition}

We remark that the list-decodability property is standard in literature.
Moreover, while the notion of average-radius list-decodability is formally
introduced by this paper, it is already implicit
in~\cite{blinovsky,blin-book,GV}. The following proposition asserts that this
is a syntactically stronger property than standard list-decodability:

\begin{proposition}
If $C$ is $(p, L)$ average-radius list-decodable, then $C$ is $(p, L)$
list-decodable.
\end{proposition}
\begin{proof}
The claim follows from the observation that the maximum distance of a list
from a center $x$ always dominates its average distance from $x$.
\end{proof}

In particular, any limitation we establish for list-decodable codes
also carries over for average-radius list-decodable codes.

Following (and extending) the notation in \cite{GHK11}, we make the following
definitions to quantify the tradeoffs in the different parameters of a code:
the rate $R$, the error-correction radius $p$, the list-size $L$, and the
weight $\lambda$ of the codewords (for ``constant weight'' codes). Further,
for general codes (without the constant-weight restriction), it is usually
more convenient to replace the rate $R$ by the parameter
$\gamma := 1- \h(p) - R$; this measures the ``gap'' to the ``limiting rate''
or the ``capacity'' of $1-\h(p)$ for $(p, O(1))$ list-decodable codes.

Fix $p, \lambda \in (0, \frac{1}{2}]$ such that $p < \lambda$, $0 \leq R \leq
1$, and a positive integer $L$.
\begin{definition} \label{def::paramzoo}
\begin{enumerate}
\itemsep=0ex

\item Say that the triple $(p, L; R)$ is {\em achievable for list-decodable
    codes} if there exist $(p, L)$ list-decodable codes of rate $R$ for
    infinitely many lengths $n$.

    Define $R_{p, L}$ to be the supremum over $R$ such that $(p, L; R)$ is
    achievable for list-decodable codes, and define
    $\gamma_{p, L} := 1 - \h(p) - R_{p, L}$. Similarly, define
    $L_{p, \gamma}$ to be the least integer $L$ such that
    $(p, L; 1- \h(p)- \gamma)$ is achievable.

\item {(\bf For constant weight codes.)} Say that the $4$-tuple
    $(\lambda; p, L; R)$ is achievable if there exists $(\lambda; p, L)$
    list-decodable codes of rate $R$. Define $R_{p, L} (\lambda)$ to be the
    supremum rate $R$ for which the $4$-tuple $(\lambda; p, L; R)$ is
    achievable.
\end{enumerate}
\end{definition}

We can also define analogous quantities for average-radius list-decoding
(denoted by a superscript $\sfavg$), but to prevent notational
clutter, we will not explicitly do so. Throughout this paper, $p$ is treated
as a fixed constant in $(0, \frac{1}{2})$, and we will not attempt to optimize
the dependence of our bounds on $p$.

% ==========

\subsection{Standard distributions and functions}

In this paper, we use `$\log$' for logarithms to base $2$ and `$\ln$' for
natural logarithms. Also, to avoid cumbersome notation, we often denote $b^z$
by $\exp_b(z)$. Standard asymptotic notations ($O$, $o$, and $\Omega$) is
employed throughout this paper; we sometimes subscript this notation by
a parameter (typically $p$) to mean that the hidden constant could depend
arbitrarily on the parameter.

Our proofs make a heavy of {\em hypergeometric distributions}, which we review
here for the sake of completeness as well as to set the notation. Suppose a
set contains $n$ objects, exactly $m < n$ of which are {\em m\/}arked, and
suppose we {\em s\/}ample $s < n$ objects uniformly at random from the set
{\em without replacement}. Let the random variable $T$ count the number of
marked objects in the sample; then $T$ follows the hypergeometric distribution
with parameters $(n, m, s)$. A simple counting argument shows that, for
$t \leq \min \{ m, s \}$,
\[ \Pr [T = t] = \frac{\binom{m}{t} \binom{n-m}{s-t} }{\binom{n}{s}} .\]
We will denote the above expression by $f(n, m, s, t)$. By convention,
$f(n, m, s, t)$ is set to $0$ if $n < \max \{ m, s \}$ or
$t > \min \{ m, s\}$.

Our proofs rely on the following two properties of hypergeometric
random variables. While these claims are standard, we have included a proof
in Appendix~\ref{apdx::subsec::hypergeo} for completeness.

\begin{fact} [Interchange property] \label{fact::hypergeosymm}
For all integers $n, m, s$ with $n \geq \max \{ m, s \}$, the hypergeometric
distribution with parameters $(n, m, s)$ is identical to that with parameters
$(n, s, m)$. That is, for all $t$, we have $f(n, m, s, t) = f(n, s, m, t)$.
\end{fact}

\begin{fact} \label{fact::hypergeostochdom}
Suppose $n, m, m', s$ are integers such that $m \geq m'$ and
$n \geq \max \{ m, s\}$. Then the hypergeometric distribution with parameters
$(n, m, s)$ {\em stochastically dominates} the hypergeometric distribution
with parameters $(n, m', s)$. That is, for all $\tau$, we have
\[
    \sum_{t = \tau}^{\infty} f(n, m, s, t) \geq
        \sum_{t = \tau}^{\infty} f(n, m', s, t).
\]
\end{fact}

Throughout this paper, we are especially concerned with the asymptotic
behaviour of binomial coefficients, which is characterized in terms of the
binary entropy function, defined as $\h(z) := -z \log z - (1-z) \log (1-z)$.
We will use the following standard estimate without proof.
\begin{fact} \label{fact::binomasymp}
Fix $z \in (0, 1)$, and suppose $n \to \infty$ such that $zn$ is an integer.
Then
\[
2^{\h(z) n - o(n)}
\leq \binom{n}{z n} \leq  \sum_{i=0}^{zn} \binom{n}{i}
\leq 2^{\h(z) n}
.\]
\end{fact}

% =========================

\section{Bounds for average-radius list-decodability} \label{sec::strongld}

In this section, we prove that the largest asymptotic rate of
$(p, L)$ average-radius list-decodable binary codes is bounded by
\[
1 - \h(p) - \frac{1}{L} - o(1) \leq R_{p, L}
    \leq 1 - \h(p) - \frac{a_p}{L^2} + o(1),
\]
where $a_p$ is a constant depending only on $p$. (Here $p$ is a fixed constant
bounded away from $0$ and $\frac{1}{2}$.) Note that the corresponding upper
and lower bounds on $\gamma := 1 - \h(p) - R$ are polynomially related,
ignoring the dependence on $p$.

We first state the rate lower bound.
\begin{theorem} \label{thm::strong-ld-pos}
Fix $p \in (0, \frac{1}{2})$ and a positive integer $L$. Then, for all $\eps
> 0$ and all sufficiently large lengths $n$, there exists a $(p, L)$
average-radius list-decodable code of rate at least $1 - \h(p)- 1/L - \eps$.
\end{theorem}

\begin{proof}
We will show that a random code of the desired rate is $(p, L)$ average-radius
list-decodable w.h.p. Consider a random code $C : \zo^{Rn} \to \zo^n$ of rate
$R := 1 - \h(p) - 1/L -\eps$; i.e., for each $x \in \zo^{Rn}$, we pick $C(x)$
independently and uniformly at random from $\zo^n$. For any $a \in \zo^n$ and
any distinct $L$-tuple $\{x_1, \ldots, x_L\} \subseteq \zo^{Rn}$, we are
interested in bounding the probability of the event that $D \leq L pn$,
where $D := \sum_{i=1}^L \dist (a, C(x_i))$.

To estimate this probability, let $A$ be the $\zo$-string of length $L n$
obtained by concatenating $a$ repeatedly $L$ times. Similarly, let $Y$ be the
$\zo$-string obtained by concatenating $C(x_1), \ldots, C(x_L)$. In this
notation, note that $D$ is simply the Hamming distance between $A$ and $Y$.
Now, $Y$ is distributed uniformly at random in $\zo^{L n}$ independently of
the choice of $A$, hence the probability that $D \leq pL n$ is at most
$\exp_2 \left( (\h(p) - 1)L n \right)$ (Fact~\ref{fact::binomasymp}).

Finally, by a union bound over the choice of $a$ and $\{ x_1, \ldots, x_L \}$,
the code fails to be $(p, L)$ average-radius list-decodable with probability
at most
\[ 2^n \binom{2^{Rn}}{L} \cdot \exp_2 \left( (\h(p)-1)L n \right) \leq
\exp_2 \left( n + (R + \h(p) - 1 ) Ln \right) =
\exp_2 \left(- \eps L n \right) ,\]
for the given choice of $R$, establishing the claim.
\end{proof}

We now show an upper bound of $1- \h(p) - a_p/L^2$ on the rate of a $(p, L)$
average-radius list-decodable code. As stated in the Introduction, the main
idea behind the construction is that instead of picking the ``bad list
decoding center'' $x$ uniformly at random, we pick it randomly {\em very close
to a designated codeword} $c^\ast$ (which itself is a uniformly random element
from $C$). Now as long as we are guaranteed to find a list of $L - 1$ other
codewords near the center, we can include $c^\ast$ in our list to lower its
average radius.

However formalizing the above intuition into a
proof is nontrivial, since our restriction of the center $x$ to be very close
to $c^\ast$ introduces statistical dependencies while analyzing the number of
codewords near $x$. We are able to control these dependencies, but this
requires some heavy calculations involving hypergeometric distributions
and the entropy function.

We are now ready to state our main result establishing a rate upper bound for
$(p, L)$ average-radius list-decodable codes. In fact, the bulk of the work is
to show an analogous upper bound for the special case of a constant-weight
code $C$, i.e., all codewords have weight exactly $\lambda n$, for some
$\lambda \in (p, \frac{1}{2})$. We can then map this bound for general codes
using a standard argument (given in Lemma~\ref{lem::fishnet}).

\begin{theorem} [Main theorem] \label{thm::strongld-neg}
Fix $p \in (0, \frac{1}{2})$, and let $L$ be a sufficiently large positive
integer. Then there exist $a_p, a_p' > 0$ (depending only on $p$) such that
the following holds (for sufficiently large lengths $n$):

\begin{enumerate}
\itemsep=0ex
\vspace{-1ex}

\item If $C$ is a $(p, L)$ average-radius list-decodable code, then $C$ has
    rate at most $1 - \h(p) - a_p / L^2 + o(1)$.

\item For $\lambda := p + a_p'/L$, if $C$ is a $(\lambda; p, L)$
    average-radius list-decodable code, then $C$ has rate at most $\h(\lambda)
    -\h(p) - a_p / L^2 + o(1)$.
\end{enumerate}
\end{theorem}

As already mentioned in Section~\ref{subsec::otherresults}, the second claim
readily implies the first via the following well-known argument (a partial
converse to this statement for list-decoding will be given in
Section~\ref{sec::constwt}):
\begin{lemma} \label{lem::fishnet}
Let $\lambda \in (p, \frac{1}{2}]$ be such that $\lambda n$ is an integer. If
$C$ is a $(p, L)$ average-radius list-decodable code of rate $R = 1 - \h(p)
-\gamma$, then there exists a $(\lambda; p, L)$ average-radius list-decodable
code of rate at least $\h(\lambda) - \h(p) - \gamma- o(1)$.
\end{lemma}
\begin{proof}
For a random center $x$, the expected number of codewords $c \in C$ with
$\dist (x, c) = \lambda n$ is exactly $|C| \cdot \binom{n}{\lambda n} 2^{-n}$.
For the assumed value of rate $R$, using Fact~\ref{fact::binomasymp}, this is
at least
\[ \exp_2 \left( (\h(\lambda)- \h(p) - \gamma - o(1)) n \right) .\]
Then there exists an $x$ such that the subcode $C' \subseteq C$ consisting of
all codewords at a distance $\lambda n$ from $x$ has rate at least
$\h(\lambda)-\h(p)-\gamma -o(1)$. The claim follows by translating $C'$ by
$-x$.
\end{proof}

Before we proceed to the proof of the first part of
Theorem~\ref{thm::strongld-neg}, we will establish the following folklore
result, whose proof illustrates our idea in a simple case.

\begin{lemma}[A warm-up lemma] \label{lem::warmup}
Fix $p, \lambda$ so that $p < \lambda \leq \frac{1}{2}$. Then, if $C$ is a
$(\lambda; p, L)$ list-decodable code, then $C$ has rate at most
$\h(\lambda) -\h(p) + o(1)$.
\end{lemma}

\begin{proof}
The main idea behind the proof is that a random center of a {\em particular
weight} (carefully chosen) is close to a large number of codewords in
expectation. Pick a random subset $S \subseteq [n]$ of coordinates of size
$\alpha n$, with $\alpha := (\lambda-p)/(1-2p)$, and let $\overline{S} := [n]
\smallsetminus S$. (The motivation for this choice of $\alpha$ will be clear
shortly.) Define the center $x$ be the {\em indicator vector} of $S$; i.e.,
$\supp(x) = S$.

Consider the set $\List$ of codewords $c \in C$ such that $\wt(c|_S) \geq
(1-p) \alpha n$; this is our candidate bad list of codewords. Then each $c \in
\List$ is close to $c$:
\[
\dist(x, c)
= \left( \alpha n - \wt (c |_{S}) \right) + \wt ( c |_{\overline{S}} )
\leq \alpha p n + (\lambda - \alpha (1-p)) n
= (\lambda - \alpha (1-2p)) n,
\]
which equals $pn$ for the given choice of $\alpha$. Hence the size of $\List$
is a lower bound on the list-size of the code.

We complete the proof by computing $\E~| \List |$. For any fixed $c \in C$,
the random variable $\wt(c|_S)$ follows the hypergeometric distribution with
parameters $(n, \lambda n, \alpha n)$, which is identical to the
hypergeometric distribution with parameters $(n, \alpha n, \lambda n)$ (see
Fact~\ref{fact::hypergeosymm}). Hence the probability that $c$ is included in
the list $\List$ is at least
\[
f (n, \alpha n, \lambda n, \alpha(1-p) n)
:=
\frac
{{\binom{\alpha n}{(1-p) \alpha n}}
{\binom{(1-\alpha) n}{(\lambda - \alpha(1-p))n}}}
{\binom{n}{\lambda n}}
=
\frac { \binom{\alpha n}{p \alpha n} \binom{(1-\alpha) n}{ p (1-\alpha) n } }
{ \binom{n}{\lambda n} }
,\]
where the second step uses the identity $\lambda -(1-p) \alpha = p(1-\alpha)$,
which holds for our particular choice of $\alpha$. As $n \to \infty$, this is
equal to
\[ \exp_2 \left( \alpha n \h(p) + (1-\alpha) n \h(p)
    - \h(\lambda) n - o(n) \right) = \exp_2((\h(p) - \h(\lambda) - o(1)) n)
.\]

Thus, by linearity of expectations, the expected size of $\List$ is at least
$|C| \cdot  \exp_2 ((\h(p) - \h(\lambda) - o(1))n)$. On the other hand, the
$(p, L)$ list-decodability of $C$ says that $|\List| \leq L$ (with probability
$1$). Comparing these lower and upper bounds on $\E~|\List|$ yields the claim.
\end{proof}

\vspace{1ex}
\begin{proofof}{Theorem~\ref{thm::strongld-neg} (part 2)} At a high level, we
proceed as in the proof of Lemma~\ref{lem::warmup}, but in addition to the bad
list $\List$ of codewords, we will a special codeword $c^\ast \in C$ such that
$\dist (x, c^\ast)$ is {\em much smaller than the codewords in $\List$}. Then
defining $\List^\ast$ to consist of $c^\ast$ and $(L-1)$ other codewords from
$\List$, we see that the {\em average} distance of $\List^\ast$ is much
smaller than before, thus enabling us to obtain an improved rate bound.

We now provide the details. Pick a uniformly random codeword $c^\ast \in C$.
Let $S \subseteq [n]$ be a random subset of $\supp(c^\ast)$ of size $\beta n$,
where the parameter $\beta$ is chosen appropriately later\footnote{At this
point, the reader might find it useful to think of both $\lambda - p$ and
$\beta$ as $\Theta(1/L)$; roughly speaking, this setting translates to a rate
upper bound of $\h(\lambda) - \h(p) - \Omega (\beta/ L)$.}
(this plays the role of $\alpha$ in Lemma~\ref{lem::warmup}). Also, let $x$ be
the indicator vector of $S$.

As before, consider the set $\List$ of codewords $c \in C$ such that
$\wt(c|_S) \geq (1-p) |S|$. For a fixed $c \in C$, the random variable
$\wt(c|_S)$ follows the hypergeometric distribution with parameters
$(\lambda n, (\lambda - \delta)n, \beta n)$, where
$\delta = \delta(c^\ast, c)$ is defined by
$\dist (c^\ast, c) := 2 \delta n$. (Observe that the normalization ensures
that $0 \leq \delta \leq \lambda$ for all pairs $c^\ast, c \in C$.) To see
this, notice that we are sampling $\beta n$ coordinates from $\supp(c^\ast)$
without replacement, and that $\wt(c|_S)$ simply counts the number of
coordinates picked from $\supp(c^\ast) \cap \supp(c)$ (the size of this
intersection is exactly $(\lambda - \delta) n$). Thus, conditioned on
$c^\ast$, the probability that a fixed $c \in C$ is included in $\List$ is
\begin{equation} \label{eqn::defineQ}
Q(\delta) :=
\sum_{w = (1-p) \beta n}^{\beta n}
                f(\lambda n, (\lambda - \delta)n, \beta n, w) .
\end{equation}
By linearity of expectations, and taking expectations over $c^\ast$,
the expected size of $\List$ can be written as
\begin{equation} \label{eqn::expLsize}
\E_{c^\ast \in C} \left[ \sum_{c \in C} Q(\delta(c^\ast, c)) \right]
= |C| \cdot \E~Q (\delta(c^\ast, c))
,\end{equation}
where {\em both} $c^\ast$ and $c$ are picked uniformly at random from $C$. The
following lemma provides a lower bound on this expectation.

\begin{lemma} \label{lem::EQest}
For $A_1 := (1-p) \log \left( \frac{1-p}{\lambda} \right) +
p \log \left( \frac{p}{1-\lambda} \right)$ and $A_2 = \frac{2}{p^2}$, we have
\[
\E~Q(\delta(c^\ast, c)) \geq
\exp_2 \left( -(A_1 \beta + A_2 \beta^2 + o(1)) n \right) ,
\]
where the expectation is taken over pairs $c^\ast, c$ of codewords.
\end{lemma}

\begin{remark}
In the above estimate, the coefficient $A_1$ is tight for all values of $p$
and $\lambda$ (assuming $\beta \to 0$ keeping $p$ and $\lambda$ fixed), but
$A_2$ can be improved significantly. For our purposes, it suffices that $A_2$
depends on $p$ alone, and not on $\lambda$ or $\beta$.
\end{remark}

\begin{proof}
By a standard application of the Cauchy-Schwarz inequality, we can show that
$\E~\delta \leq \lambda (1- \lambda)$. To see this, let $f_j$ denote the
fraction of codewords of $C$ that have $1$ in the $j$th coordinate. The weight
constraint on the codewords implies that $\sum_{j=1}^n f_j = \lambda n$.
Therefore,
\begin{align*}
\E_{c^\ast, c}~ \left[ \dist(c^{\ast}, c) \right]
&= \sum_{j=1}^n 2f_j (1- f_j)
= 2 \sum_{j=1}^n f_j - 2 \sum_{j=1}^n f_j^2
\\ &\leq  2 \sum_{j=1}^n f_j - \frac{2}{n} \left( \sum_{j=1}^n f_j \right)^2
= 2 \lambda n - 2 \lambda^2 n,
\end{align*}
and so, $\E~\delta \leq \lambda (1-\lambda)$. Now, by Markov's inequality, the
probability that $\delta \leq \lambda (1-\lambda) + 1/n$ is at least
$
    1 - \frac{\lambda (1-\lambda)}{\lambda(1-\lambda) + \frac{1}{n}} \geq
        \frac{1}{n}.
$

Moreover, using Fact~\ref{fact::hypergeostochdom} (with $\tau := \beta
(1-p)$), we know that $Q(\delta)$ is a monotonically decreasing function of
$\delta$. Therefore,
\begin{align*}
    \E~Q(\delta(c^\ast, c))
    &\geq \frac{1}{n} \cdot Q(\lambda (1-\lambda) + o(1))
    \\ &\geq \frac{1}{n} \cdot
        f \left( \lambda n, (\lambda^2 - o(1))n, \beta n, \beta(1-p) n \right)
.\end{align*}
The rest of the proof consists of lower bounding the right hand side.
As $n \to \infty$, using Fact~\ref{fact::binomasymp}, we get $\E~Q(\delta)
\geq \exp_2 (\eps n - o(n))$, where
\begin{align*}
\eps
:=
 \lambda^2 \cdot \h \left( \frac{(1-p) \beta}{\lambda^2} \right)
+ \lambda (1-\lambda)  \cdot
        \h\left( \frac{p \beta}{\lambda (1-\lambda)} \right)
- \lambda \cdot \h\left( \frac{\beta}{\lambda} \right)
.\end{align*}
We are interested in lower bounding the exponent $\eps$, and we do this by
bounding each of the above entropy terms individually using
Fact~\ref{fact::htaylor} (see Appendix~\ref{apdx::subsec::entro}), and
canceling common terms. We just mention the final bound ignoring the
intermediate steps:
\[
    \eps \geq
        \beta \left( (1-p) \log \frac{\lambda^2}{1-p} +
            p  \log \frac{\lambda(1-\lambda)}{p} - \log \lambda \right)
        - \beta^2  (\log e) \left(\frac{(1-p)^2}{\lambda^2} +
            \frac{p^2}{\lambda(1-\lambda)} \right)
.\]
Noting that
\begin{align*}
(1-p) \log \frac{\lambda^2}{1-p} +
        p  \log \frac{\lambda(1-\lambda)}{p} - \log \lambda
&= (1-p) \log \frac{\lambda}{1-p} +
        p  \log \frac{1-\lambda}{p}
= -A_1
,\end{align*}
and
\begin{align*}
(\log e) \left(\frac{(1-p)^2}{\lambda^2} +
            \frac{p^2}{\lambda(1-\lambda)} \right)
&\leq (\log e) \left( \frac{1}{p^2} + 1 \right) \leq \frac{2}{p^2}
,\end{align*}
we get the claim.
\end{proof}

We now return to the proof of Theorem~\ref{thm::strongld-neg}. From
(\ref{eqn::expLsize}) and Lemma~\ref{lem::EQest}, if the code $C$ has rate
at least $A_1 \beta + A_2 \beta^2 +o(1)$
(for a suitable $o(1)$ term), the list $\List$ has size at least $L$ in
expectation. Fix some choice of $c^\ast$ and $S$ such that $|\List| \geq L$.
Let $\List^\ast$ be any list containing $c^\ast$ and $L-1$ other codewords
from $\List$; we are interested in $\Davg (x, \List^\ast)$. Clearly, $\dist
(x, c^\ast) = (\lambda - \beta) n$. On the other hand, for $c \in \List^\ast
\smallsetminus \{ c^\ast \}$, we can bound its distance from $x$ as:
$\dist (x, c) \leq \beta p n + (\lambda - \beta (1-p)) n =
(\lambda - \beta (1-2p)) n$, where the two terms are respectively the
contribution by $S$ and $[n] \smallsetminus S$. Averaging these $L$ distances,
we get that
\[ \Davg (x, \List^\ast) \leq \left( \lambda - \beta (1-2p + 2p /
L) \right)n .\] Now, we pick $\beta$ so that this expression is at most $pn$;
i.e., set
\begin{equation} \label{eqn::betachoice}
\beta := \frac{ \lambda - p }{ 1-2p+2p/L } .
\end{equation}
(Compare with the choice of $\alpha$ in Lemma~\ref{lem::warmup}.) For this
choice of $\beta$, the list $\List^\ast$ violates the average-radius
list-decodability property of $C$.

Thus the rate of a $(p, L)$ average-radius list-decodable code is upper
bounded by $R \leq A_1\beta + A_2 \beta^2+o(1)$, where $\beta$ is given by
(\ref{eqn::betachoice}). Further technical manipulations brings this to the
following more convenient form: If $L > \frac{2p}{1-2p}$, then
\[
R \leq \left( \h(\lambda) - \h(p) \right) - \frac{B_1 (\lambda-p)}{L}
+ B_2 (\lambda - p)^2 + o(1)
.\]
for $B_1 := \frac{1}{2} p$ and $B_2 := \frac{3}{p^2 (1-2p)^2}$; see
Lemma~\ref{lem::manip-rateub} in Appendix~\ref{apdx::subsec::entro}
for a proof. Note that the second term dominates the third whenever $\lambda
-p$ is small enough. In particular, for
\[
\lambda := p + \frac{B_1}{2 B_2 L} = p + \frac{p^3 (1-2p)^2 }{12L} ,\]
the rate is upper bounded by
\[
R \leq \h(\lambda) - \h(p) - \frac{B_1^2}{4 B_2 L^2} + o(1)
= \h(\lambda) - \h(p) - \frac{p^4 (1-2p)^2}{48 L^2} + o(1)
.\]
\end{proofof}

% =========================

\section{Bounds for (standard) list-decodability} \label{sec::std-ld}

In this section, we consider the rate vs.\ list-size tradeoff for the
traditional list-decodability notion. For the special case when the fraction
of errors is close to $\frac{1}{2}$, \cite{GV} showed that any code family of
growing size correcting up to $\frac{1}{2}-\eps$ fraction of errors must have
a list-size $\Omega(1/\eps^2)$, which is optimal up to constant factors.
When $p$ is bounded away from $1/2$, Blinovsky \cite{blinovsky, blin-q-ary}
gives the best known bounds on the rate of a $(p, L)$ list-decodable code.
His results imply (see \cite{atri-ieeeit} for the calculations) that any
$(p, L)$ list-decodable code of rate $1- \h(p) - \gamma$ has list-size
$L$ at least $\Omega_p (\log (1/\gamma))$. We give a short and simple proof of
this latter claim in this section.

\begin{theorem} [\cite{blinovsky, blin-q-ary}] \label{thm::Bli2}
\begin{enumerate}
\itemsep=0ex

\item Suppose $C$ is $(\lambda; p, L)$ list-decodable code with
    $\lambda = p + \frac{1}{2} p^L$. Then $|C| \leq 2L^2/p$, independent of
    its blocklength $n$. (In  particular, the rate approaches $0$ as $n \to
    \infty$.)

\item Any $(p,L)$ list-decodable code has rate at most $1-\h(p)-\Omega_p(p^L)$.

\end{enumerate}
\end{theorem}

\begin{proof}
\begin{enumerate}

\item
For the sake of contradiction, assume that $|C| > 2L^2 / p$.  Pick a random
$L$-tuple of codewords (without replacement) $\List = \{ c_1, c_2, \ldots,
c_L \}$, and let $S$ be the set of indices $i \in [n]$ such that each $c_j \in
\List$ has $1$ in the $i$th coordinate. Define $x$ to be the indicator vector
of $S$. Note that $\dist (x, c_j) = \wt(c_j) - \wt(x) = \lambda n -|S|$. So
$\Dmax(x, \List)$ is also $\lambda n - |S|$, and hence,
$\E~\Dmax (x, \List) = \lambda n - \E~|S|$. Thus to obtain a contradiction,
it suffices to show that $\E~|S| \geq (\lambda - p)n = \frac{1}{2} p^L n$.

Let $M := |C|$ be the total number of codewords of $C$, and let $M_i$ be the
number of codewords of $C$ with $1$ in the $i^{\text{th}}$ position. Then the
probability that $i \in S$ is equal to $g(M_i) / \binom{M}{L}$, where the
function $g : \Reals^{\geq 0} \to \Reals^{\geq 0}$ is defined by $g(z) :=
\binom{\max \{ z, L-1 \}}{L}$. By standard closure properties of convex
functions, $g$ is convex on $\Reals$. (Specifically, $z \mapsto
\max \{ z, L-1 \}$ is convex over $\Reals$, and restricted to its image,
namely, the interval $[L-1, \infty)$, the function $z \mapsto \binom{z}{L}$ is
convex. Hence their composition, namely $g$, is convex as well.)

We are now ready to bound $\E~|S|$:
\[
\frac{1}{n} \E~|S|
\stackrel{(a)}{=} \frac{1}{\binom{M}{L}} \cdot \frac{1}{n}\sum_{i = 1}^{n} g(M_i)
\stackrel{(b)}{\geq} \frac{1}{\binom{M}{L}} \cdot g\left( \frac{1}{n} \sum_{i=1}^{n} M_i \right)
= \frac{g(\lambda M)}{\binom{M}{L}}
\stackrel{(c)}{=} \frac{\binom{\lambda M}{L}}{\binom{M}{L}} .
\]
Here we have used (a) the linearity of expectations, (b) Jensen's inequality,
and (c) the fact that $\lambda M \geq 2L^2 \geq L-1$. We complete the proof
using a straightforward approximation of the binomial coefficients.
\[
\frac{1}{n} \E~|S|  \geq \frac{(\lambda M - L)^L}{M^L} =
\lambda^L \left( 1 - \frac{L}{\lambda M} \right)^L
\geq \lambda^L \left( 1 - \frac{L^2}{\lambda M} \right)
\geq \frac12 \lambda^L \geq \frac12 p^L
.\]

\item
By Lemma~\ref{lem::fishnet}, the rate of a {\em general} $(p, L)$
list-decodable code is upper bounded by $1 - \h \left( p + \frac{1}{2} p^L
\right) + o(1)$, which, by Fact~\ref{fact::hsecant} (see
Section~\ref{apdx::sec::prelim} in the Appendix), is at most
$1 -  h(p) - \frac{1}{4} (1- 2p) \cdot p^L + o(1)$.
\end{enumerate}

\end{proof}

The above method can be adapted for $q$-ary codes with an additional trick:

\begin{theorem} \label{thm::Bliq}
\begin{enumerate}
\itemsep=0ex

\item Suppose $C$ is a $q$-ary $(\lambda; p, L)$ list-decodable code with
    $\lambda = p + \frac1{2L} p^L$. Then $|C| \leq 2L^2/\lambda$.

\item Suppose $C$ is a $q$-ary $(p, L)$ list-decodable code. Then there exists
    a constant $b = b_{p,q} > 0$ such that the rate of $C$ is at most
    $1- \h_q (p) - \Omega_{q, p} \left( \frac{1}{L} p^L \right)$.
\end{enumerate}
\end{theorem}

Our proof of this theorem uses the following lemma due to Erd\" os (see
Section~2.1 of \cite{jukna} for a reference.) This result was implicitly
established in our proof of Theorem~\ref{thm::Bli2}, so we will omit a formal
proof.

\begin{lemma} [Erd\"os 1964] \label{lem::Erdos}
Suppose ${\cal A}$ is a set system over the ground set $[n]$, such that each
$A \in {\cal A}$ has size at least $\lambda n$. Then, if $|{\cal A}| \geq
2L^2/\lambda$, then there exist distinct $A_1, A_2, \ldots, A_L$ in ${\cal A}$
such that $\bigcap_{i=1}^{L} A_i$ has size at least $\frac12 n \lambda^L$.
\end{lemma}

\begin{proofof}{Theorem~\ref{thm::Bliq}}
\begin{enumerate}
\item Towards a contradiction, assume $|C| \geq 2L^2/\lambda$. Consider the
    set system ${\cal A} := \{ \supp(c) ~:~ c \in C \}$. By
    Lemma~\ref{lem::Erdos}, there exists an $L$-tuple
    $\{ c_1, c_2, \ldots, c_L \}$ of codewords such that the intersection of
    their support, say $S$, has size at least $\frac12 n \lambda^L \geq
    \frac12 n p^L$. Arbitrarily partition the coordinates in $S$ into
    $L$ parts, say $S_1, \ldots, S_L$ so that each $S_j$ has size at
    least $\frac{1}{2L} p^L  n$.

    Now consider a center $x$ such that $x$ agrees with $c_j$ on all
    coordinates $i \in S_j$; for $i \notin S$, set $x_i$ to be zero. Then,
    clearly, $\dist (x, c_j) \leq \wt(c_j) - |S_j| = \lambda n -
    \frac{1}{2L} p^L \cdot n = pn$.
    Thus the list $\{ c_1, \ldots, c_L \}$ contradicts the $(p, L)$
    list-decodability of $C$.

\item From a $q$-ary generalization of Lemma~\ref{lem::fishnet} (proof
    omitted), the rate of a $(p, L)$ $q$-ary list-decodable code is at least
    $1 - \h_q \left(p + \frac{1}{2L} p^L \right)$. For $L$ large enough, this
    is at most $1 - \h_q (p) - \Omega_{q, p} \left(\frac{1}{2L} p^L\right)$,
    which implies the claim.

\end{enumerate}
\end{proofof}

% =========================

\section{Constant-weight vs.\ General codes} \label{sec::constwt}

In this section, we will understand the rate vs.\ list-size trade-offs for
constant-weight codes, that is, codes with every codeword having weight
$\lambda n$, where $\lambda \in (p, \frac{1}{2}]$ is a parameter. (Setting
$\lambda = \frac{1}{2}$ roughly corresponds to arbitrary codes having no
weight
restrictions.) As observed earlier, a typical approach in coding theory to
establish rate upper bounds is to study the problem under the above
constant-weight restriction. One then proceeds to show a strong negative
result of the flavor that a code with the stated properties must have a
constant size (and in particular {\em zero} rate). For instance, the first
part of Theorem~\ref{thm::Bli2} above is of this form. Finally, mapping this
bound to arbitrary codes, one obtains a rate upper bound of $1-\h(\lambda)$
for the original problem. (Note that Lemma~\ref{lem::fishnet} provides a
particular formal example of the last step.)

In particular, Blinovsky's rate upper bound (Theorem~\ref{thm::Bli2}) of
$1- \h(p)- 2^{-O(L)}$ for $(p, L)$ list-decodable codes follows this
approach.\footnote{For notational ease, we suppress the dependence on $p$ in
the $O$
and $\Omega$ notations in this informal discussion.} More precisely, he
proves that, under the weight-$\lambda$ restriction, such code must have zero
rate for all $\lambda \leq p + 2^{-b_p L}$ for some $b_p < \infty$. One may
then
imagine improving the rate upper bound to $1 -\h(p) - L^{-O(1)}$ {\em simply
by} establishing the latter result for correspondingly higher values of
$\lambda$ (i.e., up to $p + L^{-O(1)}$). We show that this approach cannot
work by establishing that (average-radius) list-decodable codes of positive
(though possibly small) rates exist as long as $\lambda-p \geq 2^{-O(L)}$.
Thus Blinovsky's result identifies the correct {\em zero-rate regime} for the
list-decoding problem; in particular, his bound is also the best possible if
we restrict ourselves to this approach. In this context, it is also worth
noting that for average-radius list-decodable codes,
Theorem~\ref{thm::strongld-neg} already provides a better rate upper bound
than what the zero-rate regime indicates, thus suggesting that the ``zero-rate
regime barrier'' is not an inherent obstacle, but more a limitation of the
current proof techniques.

In the opposite direction, we show that the task of establishing rate upper
bounds for constant weight codes is not significantly harder than the general
problem. Formally, we state that that if the ``gap to list-decoding capacity''
for general codes is $\gamma$, then the gap to capacity for weight-$\lambda n$
codes is {\em at least}
$\left( \frac{\lambda - p}{\frac{1}{2} - p} \right) \gamma$. Stated
differently, if our goal is to establish a $L^{-O(1)}$ lower bound on the gap
$\gamma$, then we do not lose by first passing to a suitable $\lambda$ (that
is not too close to $p$).

% ===============

\subsection{Zero-rate regime}

\begin{theorem} \label{thm::zrate}
Fix $p \in (0, \frac{1}{2})$, and set $b = b_p := \frac{1}{2}
\left( \frac{1}{2} - p \right)^2$. Then for
$L \geq \frac{1}{2b} \log \left( \frac{32}{b} \right)$ and all sufficiently
large $n$, there
exists a $(\lambda; p, L)$ average-radius list-decodable code of rate at least
$R-o(1)$, with $\lambda \leq p + 5 e^{-b L}$ and $R := \min
\{ e^{-2bL}, e^{-bL} / (6L) \} = \Omega_{p, L}(1)$.
\end{theorem}

\begin{proof}
The basic idea of the proof is that a random code is $(p, L)$
average-radius list-decodable, even if the codewords are biased to have weight
close to $pn$. We then use expurgation to ensure that all codewords have the
same weight. We now provide the details. Set $\eps := e^{-bL}$ and
$\lambda' := p+4\eps$;
verify that for the assumed values of $L$, we have $\frac{1}{2} - \lambda'
\geq \frac12 \left( \frac{1}{2}-p \right)$. Choose a random code $C :
\zo^{R n} \to \zo^n$ in the following way. For each $x \in \zo^{R n}$, each
coordinate of $C(x)$ chosen independently to be $1$ with probability
$\lambda'$ (and $0$ with the complementary probability).

Firstly, for a fixed $x \in \zo^{Rn}$, by Chernoff bound, its encoding
$C(x)$ has weight in the interval $(\lambda' \pm \eps) n$ with probability at
least $1- 2 \exp (-2 \eps^2 n) \geq 1 - \exp_2 (-2 \eps^2 n + o(n))$. By union
bound, this holds for all $x$ with probability at least $1 - \exp_2 (Rn
- 2 \eps^2 n + o(n))$.

Next, we consider the event that $C$ is $(p, L)$ average-radius
list-decodable. Specifically, we require that for every $L$-tuple of messages
$X := \{ x_1, \ldots, x_L \} \subseteq \zo^{Rn}$ and every center $a \in
\zo^n$, the encodings of the $x_i$s are $pn$-far from $a$ on average. It is
easy to bound the probability of this event for a {\em fixed} pair $(a, X)$,
and naively, we might hope to achieve this for {\em all} such pairs by a
simple union bound. However, this does not quite work, since the union bound
over $a$ contributes a $2^n$ factor loss to the probability and results in a
trivial bound. To get around this issue, we note that for any list of messages
$X$, it suffices to control the above event for a specific choice of $a$,
namely, an arbitrary {\em centroid} of the encodings of $x_1, \ldots, x_L$; we
then finish the argument by a union bound over all $X$. Since the centroid
minimizes the average distance of a center to a given list (see
Fact~\ref{fact::centroid}),
the code is now guaranteed to be $(p, L)$ average-radius list-decodable.

Fix an $L$-list
$X := \{ x_1, \ldots, x_L \}$ of messages, let $a$ denote the centroid of
their encodings. For a fixed $j \in [n]$, by Chernoff bound, the probability
that the $j^{\rm th}$ entry of $a$ is $1$ is
at most $\exp_2 \left( -2 \left( \frac{1}{2} - \lambda' \right)^2 L \right)$,
which is at most
\[ \exp_2 \left( - \frac{1}{2} \left( \frac{1}{2} - p \right)^2 L \right)
= \exp(-bL) = \eps  .\]
Moreover, the entries of $a$ in the $n$ coordinates are all independent, and
hence, by another application of the Chernoff bound (in the
multiplicative form), the weight of $a$ is at most $2\eps n$, except with
probability at most $\exp_2 (-\eps n/3)$. Assuming that this event holds, for
each $x \in X$,
\[
\dist(a, x) \geq \wt(x) - \wt(a) \geq (\lambda' - \eps) n - 2\eps n
> (\lambda' - 4 \eps) n =: pn,
\]
and hence the average distance of $X$ from $a$ is also more than $pn$.
Finally, by a union bound over $X$, we can conclude that the code is $(p, L)$
average-radius list-decodable, except with probability
$\exp_2(RLn - \eps n /3)$.

Thus, for $R = \min \{ \eps^2, \eps / (6 L) \}$, with probability
$1-o(1)$, the random code $C$ satisfies the following:

\begin{itemize}
\item Each codeword in $C$ has weight at most $(\lambda' + \eps)n$. Note that
    $\lambda' + \eps = p + 5 \eps = p + 5 e^{-bL}$.

\item $C$ is $(p, L)$ average-radius list-decodable.
\end{itemize}

Fix any $C$ with the above properties. This satisfies all our requirements,
except that its codewords could have varying weights. Fortunately, however,
this is easily fixed, since, by the pigeonhole principle, $C$ contains a
constant-weight subcode $C'$ of size at least $|C|/(n+1)$, and hence of rate
$R - o(1)$. Now, if $w_0$ denotes the weight of the codewords of $C'$, then
note that $w_0 \leq (p + 5 e^{-bL}) n$, establishing the claim with
$\lambda := w_0 / n$.
\end{proof}

Note that the {\em statement} of Theorem~\ref{thm::zrate} also yields as a
corollary $(\lambda; p, L)$ list-decodable codes of positive rate with
$\lambda$ exponentially close to $p$, since standard list-decodability is only
a weaker requirement. However, interestingly, the above {\em proof} does not
work directly because
we do not have a simple analogue of Fact~\ref{fact::centroid}
identifying the best center that minimizes the maximum radius of a list.
Indeed, the authors are not aware of any proof of this result except going
through average-radius list-decodability.

% ===============

\subsection{A reverse connection between constant-weight and arbitrary codes}

\begin{lemma} Fix $p, \lambda$ such that $0 < p < \lambda \leq \frac{1}{2}$.
Then in the notation of Definition~\ref{def::paramzoo}, if $\gamma :=
1 - \h(p) - R_{p, L}$, then
\[
\h(\lambda) - \h(p) - \gamma \leq R_{p, L}(\lambda) \leq
\h(\lambda) - \h(p) - \left( \frac{\lambda - p}{\frac{1}{2} - p} \right)
\gamma.
\]
\end{lemma}

\begin{proof}
The left inequality is essentially the content of Lemma~\ref{lem::fishnet}; we
show the second inequality here. The manipulations in this proof are of a
similar flavor to those in Lemma~\ref{lem::warmup}, but the exact details are
different.

Suppose $C$ is a $(\lambda; p, L)$ list-decodable code of blocklength $n$ and
rate $R$, such that each codeword in $C$ has weight exactly $\lambda n$. Pick
a random subset $S \subseteq [n]$ of coordinates of size $\alpha_2 n$, with
$\alpha_2 := (\lambda - p)/(\frac{1}{2} - p)$, and let $\overline{S} := [n]
\smallsetminus S$. (Interestingly, our setting of $\alpha_2$ differs from the
parameter $\alpha$ employed in the proof of Lemma~\ref{lem::warmup} only by a
factor of $2$. The motivation for this choice of $\alpha_2$ will become clear
shortly.) Consider the subcode $C'$ consisting of codewords $c \in C$ such
that $\wt(c|_S) \geq \alpha_2 n/2$. For our choice of $\alpha_2$, one can
verify that if $c \in C'$, then $c$ has weight at most $p(1-\alpha_2) n =
p |\overline{S}|$ when restricted to $\overline{S}$ (this is the motivation
behind our choice of $\alpha_2$).

Consider the restriction of $C'$ to the coordinates in $S$, $C'|_S := \{
c|_S ~:~ c \in C' \}$. Our key observation is that $C'|_S$, as a code of
blocklength $\alpha_2 n$, is $(p, L)$ list-decodable. Suppose not. Then there
exists a center $x' \in \zo^S$ and a size-$L$ list $\List \subseteq C$ such
that $\dist (x', c|_S) \leq p \alpha_2 n$ for all $c \in \List$. Now, extend
$x'$ to $x \in \zo^n$ such that $x$ agrees with $x'$ on (the coordinates in)
$S$ and is zero on the remaining coordinates. Then $\List$ violates the
$(p, L)$ list-decodability of $C$, since for every $c \in \List$,
\[
\dist (x, c) = \dist (x', c|_S) + \wt(c|_{\setcomplement{S}}) \leq
p\alpha_2 n + p (1-\alpha_2) n = pn
.\]
Therefore, $C'|_S$ must be $(p, L)$ list-decodable, and hence, by the
hypothesis of the lemma, its size is at most
$\exp_2 ((1 - \h(p) - \gamma + o(1)) \alpha_2 n)$ with probability $1$.
(It is crucial for this proof that the blocklength of $C'$ is $\alpha_2 n$,
which is significantly smaller than $n$.)

Now, for a fixed $c \in C$, the random variable $\wt(c|_S)$ follows the
hypergeometric distribution with parameters $(n, \lambda n, \alpha_2 n)$,
which is identical to the hypergeometric distribution with parameters $(n,
\alpha_2 n, \lambda n)$. Hence, the probability that $c$ is included in $C'$
is at least
\begin{align*}
f(n, \alpha_2 n, \lambda n, \alpha_2 n/2)
& =
\frac
{\binom{\alpha_2 n}{\alpha_2 n/2}
 \binom{(1-\alpha_2)n}{(\lambda - \alpha_2/2) n} }
{ \binom{n}{\lambda n} } \\
& \stackrel{(\ast)}{=}
\frac{\binom{\alpha_2 n}{\alpha_2 n/2} \binom{(1-\alpha_2)n}{p (1 - \alpha_2) n} }{ \binom{n}{\lambda n} } \\
& \geq
\exp_2 \left( \alpha_2 n + \h(p) ( 1 - \alpha_2) n - \h(\lambda) n - o(n) \right)
.\end{align*}
In the step marked $(\ast)$, we have used the the identity $\lambda -
\alpha_2/2 = p(1-\alpha_2)$, which holds for our particular choice of
$\alpha_2$. Summing this over all $c \in C$, the expected size of
$C'|_S$ is at least
\[ \exp_2 \left( Rn + \alpha_2 n + \h(p) ( 1 - \alpha_2) n
    - \h(\lambda) n - o(n) \right)
.\]

Finally, comparing the upper and lower bound on the
expected size of $C'|_S$, we get \[ R + \alpha_2 + (1-\alpha_2) \h(p) -
\h(\lambda) - o(1) \leq (1-\h(p)-\gamma) \alpha_2 + o(1) ,\] which can be
rearranged to give
the desired bound $R \leq \h(\lambda) - \h(p) - \alpha_2 \gamma + o(1)$.
\end{proof}

% =========================

\section{List-size bounds for random codes} \label{sec::random}

In this section, we establish optimal (up to constant factors) bounds on the
list-size of random codes, both general as well as linear.\footnote{In
contrast to Sections~\ref{sec::strongld}--\ref{sec::constwt}, our results
on random codes are stated as bounds on the list-size in terms of the rate.
Recall that a rate upper bound of $1 - \h_q(p) - \Omega_{q, p}(1/L)$
corresponds to a list-size bound of $\Omega_{q, p} (1/ \gamma)$ for codes of
rate $1 - \h_q(p) - \gamma$.}
Results of this vein were already shown by Rudra for the errors
case~\cite{atri-ieeeit}, based on the large near-disjoint packings of
Hamming balls implied by Shannon's
capacity theorems. Here we give a direct proof based on the second moment
method.\footnote{We remark that the argument in \cite{atri-ieeeit} is also
based on the second moment method, but applied to a more complicated random
variable.}  In addition, our proofs extend easily to give list-size bounds for
{\em list-decodable codes for erasure channels} as well.

Throughout this section and Appendix~\ref{apdx::sec::rand}, we
work with random $q$-ary codes -- both general and linear. A {\em random
$q$-ary code} (for $q \geq 2$) is simply a random map $C : \zq^k \to \zq^n$
where, for each $x \in \zq^k$, its image $C(x)$ is picked {\em independently
and uniformly at random} from $\zq^n$. On the other hand, a {\em $q$-ary
random linear code} is a random linear map $C : \F_q^k \to \F_q^n$ obtained in
the following way. We fix an arbitrary {\em basis} (typically, but not
necessarily, the standard basis) for the vector space $\F_q^k$, and the {\em
encoding of the basis vectors} is chosen independently and uniformly at random
from $\F_q^n$; the encoding map $C$ naturally extends for all messages in
$\F_q^k$ via linearity.

% ===============

\subsection{Proof idea} \label{subsec::rand-pfidea}

Our results proceed directly via the second moment method. Towards this goal,
we define a random variable $W$ that counts the number of {\em witnesses}
(i.e., a bad list of codewords together with the center) that {\em certify the
violation of the $(p, L)$ list-decodability property}. Thus the code is
$(p, L)$ list-decodable if and only if $W = 0$. We then show that (a) $W$ has
large expectation (i.e., $\E~W$ is exponential in $n$), but (b) its variance
is relatively small (i.e., $\Var~W / (\E~W)^2$ is exponentially small in $n$).
Then using the Chebyshev inequality (Fact~\ref{fact::Chebyshev}), we can
conclude that $W > 0$, except with an exponentially small probability, which
is what we set out to show.

As a particular example, consider the case of random general codes under
errors. In this case, the ``potential violations'' of the list-decoding
property are indexed by pairs $(a, X)$, where $a \in \zo^n$ is an arbitrary
center, and  $X$ is an arbitrary distinct $L$-tuple of messages $\{ x_1, x_2,
\ldots, x_L \} \subseteq \zo^k$. We thus define the indicator random
variable $\I(a, X)$ for the event that $\dist (a, C(x)) \leq pn$ for all
$x \in X$, and let $W := \sum_{a, X} \I (a, X)$. The mean and variance
estimates for $W$ follow by standard calculations. See the formal proofs for
details.

% ===============

\subsection{Error list-decodability bounds}
\label{subsec::rerro}

We state our bounds for standard list-decodable codes (under errors),
deferring the complete proofs to
Appendices~\ref{apdx::subsec::rerro-gen}~and~\ref{apdx::subsec::rerro-lin}.

\begin{theorem} \label{thm::rerro}
Fix $q \geq 2$, $0 < p < 1 - 1/q$ and $\gamma > 0$.

\begin{enumerate}
\item \label{itm::rerro-gen}
    A random $q$-ary code of rate $1-\h_q(p)-\gamma$ is
    $\left(p, \frac{1-\h_q(p)}{2\gamma} \right)$ list-decodable with
    probability at most $\exp_q \left( -\Omega_{p, \gamma}(n)\right)$.

\item \label{itm::rerro-lin}
    A random $q$-ary {\em linear} code of rate $1-\h_q(p)-\gamma$ is
    $\left(p, \frac{\delta_{q, p}}{2\gamma} \right)$ list-decodable with
    probability at most $\exp_q \left( -\Omega_{p, \gamma}(n)\right)$. Here,
    $\delta_{q, p}$ is a constant depending on only $q$ and $p$.

\end{enumerate}
\end{theorem}

% ===============

\subsection{Erasure list-decodability bounds} \label{subsec::reras}

The technique outlined in Section~\ref{subsec::rand-pfidea} also extends
to give list-size bounds for random $q$-ary codes under the {\em erasure
model}, which we now review briefly. In this model, the output alphabet is the
usual alphabet $\zq$ augmented with a special {\em erasure symbol}
`$\erasym$'.
For a string $a \in \left( \zq \cup \{\erasym\} \right)^n$, define
$\suppstar(a)$ to
be the set of
indices $i$ such that $a_i~\ne~\erasym$. Also, we say that
$a,b \in (\zq \cup \{?\})^n$ {\em agree} with each other if $a_i = b_i$ for
all $i \in \suppstar(a) \cap \suppstar(b)$.

\begin{definition}
A code $C \subseteq \zq^n$ is said to be {\em $(p, L)$ erasure list-decodable}
if for all $a \in (\zq \cup \{?\})^n$ satisfying $|\suppstar(a)| = (1-p)n$, at
most $L-1$ codewords of $C$ (treated as strings over $(\zq \cup \{?\})$) agree
with $a$.
\end{definition}

We are now ready to state our bounds for random (general and linear) codes
under erasures. Note the exponential gap between the list-sizes of linear and
general random codes under erasures.
\begin{theorem} \label{thm::reras}
Fix $q \geq 2$, $0 < p < 1$ and $\gamma > 0$.

\begin{enumerate}

\item \label{itm::reras-gen}
    A random $q$-ary code of rate $1- p-\gamma$ is
    $\left(p, \frac{1- p}{2\gamma} \right)$ erasure list-decodable with
    probability at most $\exp_q \left( -\Omega_{p, \gamma}(n)\right)$.

\item \label{itm::reras-lin}
    Let $q$ be a prime power. A random $q$-ary {\em linear} code of rate
    $1- p -\gamma$ is
    $\left(p, \frac{1}{q} \cdot \exp_2 \left(\frac{p(1-p)}{16\gamma} \right)
    \right) $ erasure
    list-decodable with
    probability at most $\exp_2 \left( -\Omega_{p}(n)\right)$.

\end{enumerate}
\end{theorem}

The proofs for the above two bounds appear respectively in
Appendices~\ref{apdx::subsec::reras-gen}~and~\ref{apdx::subsec::reras-lin}.

\bibliographystyle{abbrv}

\appendix
\parskip=1ex

% =========================

\section{Technical results on standard functions} \label{apdx::sec::prelim}
\subsection{Properties of hypergeometric distributions}
\label{apdx::subsec::hypergeo}

\begin{proofof} {Fact~\ref{fact::hypergeosymm}}
 We consider a modification of the experiment in the definition of the
hypergeometric distribution. Consider a set of $n$ distinguishable objects
that are marked by two players, Alice and Bob. Alice picks $m$ objects
uniformly at random and marks it `A'. Simultaneously, Bob picks $s$ objects
uniformly at random and marks it `B'. Moreover, the choices of Alice and Bob
are {\em independent} of each other. We claim that the number of objects $T$
marked by {\em both} Alice and Bob follows the hypergeometric distribution
with parameters $(n, m, s)$. Indeed, {\em conditioned} on the subset $A$ of
objects selected by Alice, the number of objects {\em from $A$} that are
picked by Bob follows the hypergeometric distribution with parameters
$(n, m, s)$ (independent of $A$); we now obtain the claim by unconditioning on
$A$.

Note that the above experiment is symmetric w.r.t. Alice and Bob, and hence
the same argument shows that $T$ follows the hypergeometric distribution with
parameters $(n, s, m)$ as well. The lemma now follows.
\end{proofof}

\begin{proofof} {Fact~\ref{fact::hypergeostochdom}}
Consider an urn containing $n$ balls, of which exactly $m'$ are black,
$m - m'$ are green, and the remaining are white. Sample $s$ balls from the urn
without replacement. Then, the number $B$ of black balls picked
follows the hypergeometric distribution with parameters $(n, m', s)$, whereas
the number $N$ of nonwhite (i.e., black or green) balls picked follows the
hypergeometric distribution with parameters $(n, m, s)$. Since, for any
outcome, it holds that $N \geq B$, the probability that $N \geq \tau$ is at
least that of the event that $B \geq \tau$, which is what we wanted to show.
\end{proofof}

\begin{remark} The joint random variable $(B, N)$ is a
{\em stochastic coupling} between the two hypergeometric distributions.
\end{remark}

% ===============

\subsection{Properties of the binary entropy function}
\label{apdx::subsec::entro}

In this section, we will prove some standard properties of the binary entropy
function used in this paper.

\begin{fact} \label{fact::hsecant}
For any $p, \lambda$ such that $0 < p < \lambda \leq \frac{1}{2}$, we have
\[
\h(\lambda) - \h(p) \geq \frac{1}{2}(1 - 2p) \cdot (\lambda - p)
.\]
\end{fact}

\begin{proof} We begin with the identity
\[
\h(\lambda) - \h(p)
= \int_{p}^{\lambda} \h'(z)~dz
= (\log e) \int_p^{\lambda} \ln \left( \frac{1-z}{z} \right)~dz .
\]
For $u \geq 1$, we have $\ln u \geq \frac{u-1}{u}$, which implies that
for $0 < z \leq \frac{1}{2}$,
\[
\ln \left( \frac{1-z}{z} \right) \geq \frac{\frac{1-z}{z} - 1}{\frac{1-z}{z}}
= \frac{1-2z}{1-z} \geq (1-2z).
\]
Therefore,
\[
\h(\lambda) - \h(p) \geq (\log e) \int_p^\lambda (1-2z)~dz
= (\log e) (1 - \lambda - p) (\lambda - p)
\geq (\log e) \left(\frac{1}{2} - p \right) (\lambda - p),
\]
which establishes the claim.

\end{proof}

\begin{fact} \label{fact::htaylor}
For all $z \in (0, 1)$, we have $z \log (1/z) + (\log e) (z - z^2) \leq
\h(z) \leq z \log (1/z) + (\log e) z $.
\end{fact}

\begin{proof}
After expanding the definition of $\h(\cdot)$, the above inequality reduces to
\[
z - z^2 \leq - (1-z) \ln (1-z) \leq z .
\]
We can equivalently write this as
\[
\ln (1 - z) \leq -z, \text{ and }
\ln \left( 1 + \frac{z}{1 - z} \right) \leq \frac{z}{1-z},
\]
both of which are special cases of the standard inequality $\ln (1 + z) \leq
z$ valid for all real $z$.
\end{proof}

Next, we show how to massage the rate upper bound given in
Theorem~\ref{thm::strongld-neg} in Section~\ref{sec::strongld} into a more
convenient form. For the remainder of the section, we set
$A_1 := (1-p) \log \left( \frac{1-p}{\lambda}\right)
+ p \log \left( \frac{p}{1-\lambda} \right)$, and $A_2 := \frac{2}{p^2}$.
\begin{lemma} \label{lem::ubA}
\[
    A_1 \leq (1-2p) \cdot \frac{\h(\lambda) - \h(p)}{\lambda-p}
                + \frac{5}{p} (\lambda - p)
.\]
\end{lemma}

\begin{proof}
We begin with
\begin{align*}
A_1 &= (1-p) \log \left( \frac{1-p}{\lambda} \right) +
            p \log \left( \frac{p}{1-\lambda} \right)
\\ &\leq (1-p) \log \left( \frac{1-p}{p} \right) +
            p \log \left( \frac{p}{1-\lambda} \right)
\\ &= (1-2p) \log \left( \frac{1-p}{p} \right) +
            p \log \left( \frac{1-p}{1-\lambda} \right)
\\ &= (1-2p) \h'(p) +  p \log \left( \frac{1-p}{1-\lambda} \right)
        \label{eqn::A1ub}
.\end{align*}

To complete the proof, we bound each term separately. First,
\begin{align*}
\h'(p)
&= \h'(\lambda) - \int_p^{\lambda} \h''(z)~dz
= \h'(\lambda) + \int_p^{\lambda} \frac{\log e}{z(1-z)}~dz
\\ &\leq \h'(\lambda) + \int_p^{\lambda} \frac{4}{z}~dz
\leq \h'(\lambda) + \frac{4(\lambda - p)}{p}
.\end{align*}
Also, by the concavity of $\h$, $\h(\lambda) - \h(p) \geq
\h'(\lambda) (\lambda - p)$, so $\h'(p) \leq
\frac{\h(\lambda) - \h(p)}{\lambda - p}  + \frac{4(\lambda - p)}{p}$.
On the other hand, applying the inequality $\ln z \leq z -1$ with $z
= \frac{1-p}{1-\lambda}$, we get
\[ \log \left (\frac{1-p}{1-\lambda} \right)
        \leq (\log e) \frac{\lambda - p}{1-\lambda} \leq 4(\lambda - p)
        \leq \frac{\lambda - p}{p^2}
,\]
since $p < \frac{1}{2}$ and $e < 4$. Plugging these in the upper
bound for $A_1$ gives the claim.
\end{proof}

\begin{lemma} \label{lem::manip-rateub}
Fix $\eps \in \left(0, \frac{1 - 2p}{2p} \right)$, and set
$\beta := (\lambda - p)/(1-2p+ 2p\eps)$. Then
\[ A_1\beta  + A_2 \beta^2 \leq \h(\lambda) - \h(p) - B_1 \eps (\lambda - p)
         + B_2 (\lambda - p)^2 \]
for $B_1 := \frac{1}{2} p$ and $B_2 := \frac{3}{p^2 (1-2p)^2}$. (Note that
$B_1$ and $B_2$ are independent of $\lambda$ and $\eps$.)
\end{lemma}

\begin{proof} From Lemma~\ref{lem::ubA}, we have
\begin{align*}
A_1 \beta
&\leq
\left[ (1-2p) \cdot \frac{\h(\lambda) - \h(p)}{\lambda - p}
  + \frac{5(\lambda - p)}{p} \right] \cdot \frac{\lambda - p}{1-2p+2p \eps}
\\ &\leq
\frac{1-2p}{1-2p+2p\eps} \cdot (\h(\lambda) - \h(p))
  + \frac{5 (\lambda - p)^2}{p(1-2p)} .
\end{align*}
Assuming $0< \eps < \frac{1-2p}{2p}$, we can upper bound this by
\begin{align*}
A_1 \beta &\leq
 \frac{1-2p - p\eps}{1-2p} \cdot (\h(\lambda) - \h(p))
    +  \frac{5 (\lambda - p)^2}{p(1-2p)}
 \\ &= \h(\lambda) - \h(p) - \frac{\h(\lambda) - \h(p)}{1-2p} \cdot p\eps
    +  \frac{5 (\lambda - p)^2}{p(1-2p)}
 \\ &\leq \h(\lambda) - \h(p) - \frac{p\eps (\lambda - p)}{2}
    +  \frac{5 (\lambda - p)^2}{p(1-2p)}
\end{align*}
using Fact~\ref{fact::hsecant}. Also, $A_2 \beta^2 \leq
\frac{2(\lambda - p)^2}{p^2 (1-2p)^2}$. Thus,
\begin{align*}
A_1\beta + A_2 \beta^2
&\leq \h(\lambda) - \h(p) - \frac{p\eps (\lambda - p)}{2}
  + \frac{5(\lambda - p)^2}{p(1-2p)} + \frac{2(\lambda - p)^2}{p^2 (1-2p)^2}
\\ &\leq \h(\lambda) - \h(p) -  \frac{p\eps (\lambda - p)}{2}
  + \frac{3 (\lambda - p)^2}{p^2 (1-2p)^2}
.\end{align*}

\end{proof}

% =========================

\section{List-decoding bounds for random codes}  \label{apdx::sec::rand}

Throughout this section, we fix the parameters $q$, $p$, and $n$. For $a \in
\zq^n$, let $\ball_q (a, pn)$ be the $q$-ary Hamming ball with center $a$ and
radius $pn$.
Let $\mu$ denote the fraction of points of $\zq^n$ that are inside a Hamming
ball of radius $pn$; i.e., $\mu = |\ball_q(a, pn)| / q^n$ for an arbitrary
$a \in \zq^n$.
We need the following estimate on $\mu$ (this generalizes
Fact~\ref{fact::binomasymp} for larger alphabet sizes):

\begin{fact} \label{fact::volest}
As $n \to \infty$,
$
\exp_q ( (\h_q(p) - 1 - o(1)) n ) \leq \mu \leq \exp_q ((\h_q(p) - 1) n)
$.
\end{fact}

We also need the following simple corollary of Chebyshev's inequality:
\begin{fact} \label{fact::Chebyshev}
Let $W$ be a nonnegative random variable. Then, $W = 0$ with probability
at most $\frac{\Var~W}{(\E~W)^2}$.
\end{fact}

% ===============

\subsection{Proof of part~\ref{itm::rerro-gen} of
Theorem~\ref{thm::rerro} (random general codes under errors)}
\label{apdx::subsec::rerro-gen}

Consider a random code $C: \zq^k \to \zq^n$, where $k := (1 -
\h_q(p) - \gamma) n$. Fix a positive integer $L$, to be specified later.
For any center $a \in \zq^n$, and any (ordered) list of $L$ {\em messages}
$X := ( x_1, x_2, \ldots, x_L ) \subseteq \zq^k$, let $\I (a, X)$ be the
indicator random variable for the event that the encoding of $x$ falls inside
the ball $\ball_q(a, pn)$ for all $x \in X$. Moreover, define $W :=
\sum_{a, X} \I(a, X)$. Clearly, the code $C$ is $(p, L)$ list-decodable
if and only if $W > 0$.

For a fixed center $a$ and a fixed message $x$, the event that the encoding of
$x$ falls inside $\ball_q(a, pn)$ occurs with probability $\mu$; since the
encodings of distinct messages are statistically independent,
$\Pr~\I (a, X) = \mu^L$. Also, assuming $k \geq L+1$, the number of possible
$(a, X)$ pairs is at least
$\frac{1}{2} q^{kL} \cdot q^n$, since the number of ordered $L$-lists $X$ of
distinct messages is
\[
q^k (q^k - 1)\cdots (q^k - L + 1) \geq
q^{kL} \left( 1 - \sum_{i=0}^{L-1} \frac{i}{q^k} \right)
= q^{kL} \left( 1 - \frac{\binom{L}{2}}{q^k} \right)
\geq q^{kL} \left( 1 - \frac{2^{L}}{2^k} \right)
\geq \frac{1}{2} q^{kL}
.\]
Therefore, by linearity of expectations, $\E~W \geq \frac{1}{2} \mu^L q^n
q^{kL}$.

We now upper bound the variance of $W$. For two lists of messages $X$ and $Y$,
define the {\em intersection parameter} $l = l(X, Y) := |X \cap Y|$. If $X$
and $Y$ are disjoint (equivalently, if $l (X, Y) = 0$), then the events
$\I(a, X)$ and $\I(b, Y)$ are independent for any pair of centers $a, b$.
Therefore,
\begin{align*}
\Var~W
&=  \sum_{X, Y} \sum_{a, b} (\E~[\I(a, X) \I(b, Y)] -
    \E~[\I(a, X)] \cdot \E~[\I(b, Y)])
\\ &=  \sum_{X \cap Y \ne \emptyset} \sum_{a, b} (\E~[\I(a, X) \I(b, Y)] -
    \E~[\I(a, X)] \cdot \E~[\I(b, Y)])
\\ &\le \sum_{X \cap Y \neq \emptyset} \sum_{a, b} \E~[\I(a, X) \I(b, Y)]
\\ &= \sum_{X \cap Y \neq \emptyset} \sum_{a, b}
      \P~[ \I(a, X) = 1 \text{ and } \I(b, Y) = 1 ].
\\ &= \sum_{l=1}^L \sum_{|X \cap Y| = l} \sum_{a, b}
      \P~[ \I(a, X) = 1 \text{ and } \I(b, Y) = 1 ].
\end{align*}

For convenience, we convert the inner summation into an expectation
by randomizing over the centers $a, b$:
\begin{equation} \label{eqn::varWmaster}
\Var~W
\le q^{2n} \sum_{l=1}^L \sum_{|X \cap Y| = l}
\P_{a, b, C}~[ \I(a, X) = 1 \text{ and } \I(b, Y) = 1 ].
\end{equation}
Here, in addition to the randomness of the code, the centers $a$ and $b$ are
picked uniformly at random from $\zq^n$.

Fix $0 < l \leq L$, and a pair $(X, Y)$ such that $|X  \cap Y| = l$. Fix
an arbitrary $z \in X \cap Y$; such a $z$ is guaranteed to exist since $X$
and $Y$ intersect. Now, the event ${\cal E}$ that
$\I(a, X) = \I(b, Y) = 1$ can be equivalently expressed as the
conjunction of the events
\begin{itemize}
\item Both $a, b$ fall inside $\ball_q(C(z), pn)$;
\item For each $x \in X \smallsetminus z$, the encoding of $x$ falls inside
    $\ball_q(a, pn)$; and
\item For $y \in Y \smallsetminus X$, the encoding of $y$ falls inside
    $\ball_q(b, pn)$.
\end{itemize}
The first event occurs with probability $\mu^2$, and conditioned on the
choice of $a$ and $b$, the second and third events occur with probabilities
$\mu^{L-1}$ and $\mu^{L-l}$ respectively (and they are independent given $a$
and $b$). Therefore the probability of ${\cal E}$ is $\mu^{2L-l+1}$. Finally,
by an easy counting, the number of pairs
$(X, Y)$ with $|X \cap Y| = l$ is at most $L^{2L} q^{k(2L-l)}$. Thus, we can
bound the variance of $W$ as
\[
\Var~W
\leq q^{2n} \sum_{l = 1}^L L^{2L} q^{k(2L-l)} \mu^{2L-l+1}
.\]
Dividing by $(\E~W)^2$, we get
\[
\frac{\Var~W}{(\E~W)^2} \leq \sum_{l = 1}^L 4 L^{2L} ( q^k \mu )^{-l} \mu .
\]
For our choice of parameters, we have $q^k \mu = q^{-\gamma n}$, and hence
\[
\frac{\Var~W}{(\E~W)^2} \leq \sum_{l = 1}^L L^{4L} q^{\gamma l n} \mu
\leq L^{4L+1} q^{\gamma L n } q^{- (1 - \h_q(p))n}.
\]
This quantity is $\exp_q \left( -\Omega_{p, \gamma}(n) \right)$ for $L :=
\frac{1-\h_q(p) }{ 2\gamma }$, and hence we are done by
Fact~\ref{fact::Chebyshev}.

% ===============

\subsection{Proof of part~\ref{itm::rerro-lin} of Theorem~\ref{thm::rerro}
    (random linear codes under errors)}
        \label{apdx::subsec::rerro-lin}

We follow the same outline as in Appendix~\ref{apdx::subsec::rerro-gen}, so
we will only highlight the differences. Let $C$ be a random {\em linear} code
of blocklength $n$ and {\em dimension} $k = (1-\h_q(p)- \gamma)n$.
We consider pairs $(a, X)$ as before, but we now allow only {\em linearly
independent list of messages $X$}. Moreover, the definition of $W$ is
unchanged, except that we sum over only the admissible $X$. Finally, we
modify the definition of the parameter $l$ to take linearity into account. For
a pair of lists $X$ and $Y$ (each of which is linearly independent), we define
$l = l(X,Y) := \dim(\vecspan(X) \cap \vecspan(Y))$ (where, for any set $Z$ of
message vectors, $\vecspan(Z)$ denotes its linear span). Note that $l = 0$
if and only if they $X$ and $Y$ are linearly independent of each other.

For any {\em linearly independent set} $X$, the encodings of vectors in $X$
are statistically independent, and hence $\E~\I(a, X) = \mu^L$. Once again,
the number of linearly independent lists $X$ is again at least
$\frac{1}{2} q^{kL}$; indeed, the number of such lists is
\[
(q^k - 1) (q^k - q) \cdots (q^k - q^{L-1}) \geq
q^{kL} \left( 1 - \sum_{i=0}^{L-1} q^{i-k} \right) \geq
q^{kL} \left( 1 - q^{L-k} \right) \geq
\frac{1}{2} q^{kL}
.\]
Therefore, as before, $\E~W \geq \frac{1}{2} (q^{k} \mu)^L q^n$.

As before, the events $\I(a, X)$ and $\I(b, Y)$ are statistically
independent whenever $X$ and $Y$ are linearly independent, i.e., $l = 0$.
Therefore, as before, we can bound the variance of $W$ by
\[
    \Var~W \leq
        q^{2n} \sum_{l=1}^L \sum_{l(X, Y) = l}
            \P_{a, b, C}~[{\cal E}]
,\]
where ${\cal E}$ is the event that $\I(a, X) = 1$ and $\I(b, Y) = 1$. Now,
fix an $l$ such that $1 \leq l \leq L$, and fix a pair $X, Y$ such that
$\dim (\vecspan\ X \cap \, \vecspan\ Y) = l$. Then, $Y$ can be partitioned as
$Y = Y_0 \cup Y_1$, with (a) $|Y_0| = l$ and $|Y_1| = L-l$, (b) $X$ is
linearly independent from $Y_1$, and (c) $Y_0 \subseteq \vecspan (X \cup \
Y_1)$. Fix an arbitrary $y_0 \in Y_0$. Then, by the span condition, we can
write $y_0 = \sum_{u \in X \cup Y_1} \theta_u \cdot u$ for some set of
{\em scalars} $\{ \theta_u \}_{u \in X \cup Y_1}$. Note that it is possible
that $y_0$ lies in the span of $X$. But, since $Y$ is an independent set,
$y_0$ {\em cannot} be written as a linear combination of vectors from $Y_1$
alone; in particular, there exists some $u \in X$ with $\theta_u \neq 0$.

In order to upper bound the probability of ${\cal E}$, we estimate the
probability that $C(y_0) \in \ball_q(b, pn)$, after {\em conditioning} on
the subevent ${\cal E}'$ that $C(u) \in \ball_q(a, pn)$ for all $u \in X$, and
$C(u) \in \ball_q(b, pn)$ for all $u \in Y_1$. (It is easy to check that the
latter event occurs with probability $\mu^{|X \cup Y_1|} = \mu^{2L - l}$.)

At this point, it is convenient to re-center the vectors in $X \cup Y_1$ as
follows: For $u \in X$, define $C'(u) := C(u) - a$,
and for $u \in Y_1$, define $C'(u) := C(u) - b$. After conditioning on
${\cal E}'$, the random variables $C'(u)$ (for $u \in X \cup Y_1$) are i.i.d.\
and are uniformly distributed inside the ball $\ball_q(\zerovec, pn)$;
furthermore, they are also independent of the choice of $a$ and $b$. In terms
of these new random variables, we can write \[
C(y_0) - b =
\sum_{u \in X \cup Y_1} \theta_u \cdot C'(u)
+ \left( \sum_{u \in X} \theta_u \right) \, a
+ \left( \sum_{u \in Y_1} \theta_u - 1 \right) \, b
.\]

We claim that conditioned on ${\cal E}'$, $C(y_0) - b \in
\ball_q(\zerovec, pn)$ occurs with probability at most $q^{-\Omega(n)}$.
We discuss two cases:
\begin{enumerate}

\item Suppose $\sum_{u \in X} \theta_u \neq 0$, {\em or}  $\sum_{u \in Y_1}
    \theta_u \neq 1$. Then, conditioned on the choice of $C'(u)$s, the random
    variable $C(y_0) - b$ is distributed uniformly at random inside $\F_q^n$
    and hence falls inside $\ball_q(\zerovec, pn)$ with probability $\mu$.

\item
Suppose that $\sum_{u \in X} \theta_u = 0$, {\em and} $\sum_{u \in Y_1}
\theta_u = 1$. In this case, we have
\begin{equation} \label{eqn::ballvecsum}
    C(y_0) - b = \sum_{u \in X \cup Y_1 ~:~ \theta_u \neq 0}
                    \theta_u \cdot C'(u)
.\end{equation}
Thus, if $m := |\{ u ~:~ \theta_u \neq 0 \}|$, then $C(y_0) - b$ is a sum of
{\em $m$ points sampled independently and uniformly from the ball
$\ball_q(\zerovec, pn)$}. Also, as observed earlier, there exists some $u \in X$
such that $\theta_u \neq 0$; moreover, since $\sum_{u \in X} \theta_u = 0$,
there are at least {\em two} $u$'s in $X$ with $\theta_u \neq 0$;
i.e., $m \geq 2$. We now bound the probability of ${\cal E}$ conditioned on
${\cal E}'$ using the following fact:
\begin{lemma} \label{lem::GHK}
For every $q \geq 2$ and every $p \in (0, \frac{1}{2})$, there exists
$\delta = \delta_{q, p}$ such that the following holds all large enough integers $n$. If
$m \geq 2$, and if  $v_1, v_2 , \ldots, v_m$ are $m$ independent and uniformly
random samples from $\ball_q(\zerovec, pn)$, then the
probability that $v_1 + v_2 + \cdots + v_m  \in \ball_q(\zerovec, pn)$ is
bounded by $n^{O(m)} \cdot q^{- \delta n}$.
\end{lemma}
We skip a formal proof of this lemma. A special case of this statement
corresponding to $m=q=2$ can be found in~\cite{GHK11} (see Lemma~7), and the
proof given there generalizes to give our claim with syntactic modifications.

We now return to the proof of Theorem~\ref{thm::rerro}. Since $m \leq 2L =
O_{n \to \infty}(1)$, Lemma~\ref{lem::GHK} implies that, conditioned on
${\cal E}'$, the stated event ${\cal E}$ also occurs with probability at most
$q^{- \delta n + O(L \log n)} = q^{-\delta n + o(n)}$. (Without loss of
generality, we may choose $\delta$ small enough so that this bound is larger
than $\mu$.)
\end{enumerate}
Therefore, the conditional probability of ${\cal E}$ is at most the maximum of
the two cases, namely $\exp_q( - \delta_{q, p} n + o(n))$.
To complete the variance bound, we need an estimate on the number of pairs
$(X, Y)$ such that $l (X, Y) = l$. Partition $Y$ as $Y_0 \cup Y_1$ as before.
Now, $X \cup Y_1$ can be picked in at most $q^{k(2L - l)}$ ways. Also, for
each $y \in Y_0$, we can write $y$ as a linear combination of vectors in $X
\cup Y_1$ in at most $q^{2L-l} \leq q^{2L}$ ways. Thus the total number of
pairs $(X, Y)$ with $l(X, Y) = l$ is at most $q^{2Ll} \cdot q^{k(2L - l)}$.
Thus, the variance can be bounded as
\begin{align*}
    \Var~W
        &\leq
        q^{2n} \sum_{l=1}^{L} q^{2Ll} \cdot q^{k(2L -l)} \mu^{2L-l} q^{- \delta n + o(n)}
        \\ &\leq
        \sum_{l=1}^{L} 4 (\E~W)^2 \cdot q^{2Ll} \left( q^k \mu \right)^{-l} q^{-an + o(n)}
        \\ &\leq
        4 (\E~W)^2 \cdot  \sum_{l=1}^{L} q^{2Ll} q^{\gamma l n - an + o(n)}
        \\ &\leq
        4 L q^{2L^2} q^{\gamma L n - an + o(n)} \cdot (\E~W)^2
.\end{align*}
Therefore, as before, the probability that $W = 0$ is also at most
$\exp_q (\gamma L n - an + o(n))$. Thus, setting $L := \delta / (2 \gamma)$, the
claim follows.

% ===============

\subsection{Proof of part~\ref{itm::reras-gen} of Theorem~\ref{thm::reras}
(random general codes under erasures)}
\label{apdx::subsec::reras-gen}

Consider a random code $C : \zq^k \to \zq^n$, where $k = (1 - p - \gamma)n$.
Let ${\cal A}$ be the set of potential inputs to the decoding algorithm, that is,
${\cal A} := \{ a \in (\zq \cup \{?\})^n ~:~ |\suppstar(a)| = (1-p)n \}$.
We modify the definition of $W$ as follows. For every $a \in {\cal A}$ and
ordered $L$-list $X$ of messages, define $\I(a, X)$ to be the indicator random
variable for the event that, for all $x \in X$, the encoding $C(x)$ of $x$
agrees with $a$; finally, in the definition of $W$, we consider only
$(a, X)$ pairs of the above form. As in the errors case, the code is
$(p, L)$ erasure list-decodable if and only if $W = 0$.

For every $a \in {\cal A}$ and $x \in \zq^k$, the encoding of $x$ agrees with
$a$ with probability $q^{-(1-p)n}$, and hence by independence, the probability
that $\I(a, X) = 1$ is $\exp_q \left(-(1-p) Ln \right)$. Therefore,
\[
    \E~W \geq
        q^{- (1-p) Ln} \cdot \binom{n}{np} q^{(1-p) n} \cdot
        \frac{1}{2} q^{k L}
,\]
where the second factor is the number of possible $a$, and the third term
is a lower bound on the number of $X$'s. Moreover, proceeding as before, we
can bound the variance of $W$ by
\begin{equation} \label{eqn::varexpr-B3}
    \Var~W \leq
        \sum_{l=1}^L \sum_{|X \cap Y| = l} \sum_{a, b}
        \P~\left[ {\cal E} \right]
,\end{equation}
where ${\cal E}$ is the event that $\I(a, X) = \I(b, Y) = 1$.

Now, fix an arbitrary pair $(X, Y)$ with $|X \cap Y| = l > 0$. Observe that
the event ${\cal E}$ implies that both $a$ and $b$ agree with the encoding
$C(u)$ of some $u \in X \cap Y$ (indeed, such a $u$ is guaranteed to exist).
Since $C(u)$ is a string over $\zq$ (i.e., it does not contain any
`$\erasym$''s), it follows that $a$ and $b$ must themselves agree with each
other. Moreover, the event ${\cal E}$ requires that (a) $C(x)$ agrees with $a$
for all $x \in X \smallsetminus Y$, (b) $C(y)$ agrees with $b$ for all $y \in
Y\smallsetminus X$, and (c) $C(z)$ agrees with {\em both} $a$ and $b$ for
$z \in X \cap Y$. Therefore, the probability of ${\cal E}$ is at most
\[
    \exp_q \left(- |S||X \smallsetminus Y| - |T||Y \smallsetminus X|
        -|S\cup T||X \cap Y| \right)
        = \exp_q \left( -2 (1-p) (L-l) n -|S\cup T| l \right)
,\]
where $S := \suppstar(a)$ and $T := \suppstar(b)$. Now, for a given pair $(S,
T)$, the number of pairs of centers $(a, b)$ such that (a) $\suppstar(a) = S$,
(b) $\suppstar(b) = T$, and (c) $a$ and $b$ agree with each other (i.e.,
$a|_{S \cap T} = b|_{S \cap T}$), is equal to $q^{|S \cup T|}$. Thus, the
inner summation in (\ref{eqn::varexpr-B3}),
\begin{align*}
\sum_{a, b} \P~\left[ \I(a, X) = 1 \text{ and } \I(b, Y) = 1 \right]
&= \sum_{S, T} \exp_q \left( -2(1-p) (L-l)n - |S \cup T| (l-1) \right)
\\ &\leq \binom{n}{p n}^2 \exp_q \left( -2(1-p) (L-l) n -
        (1-p) n (l-1) \right)
\\ &= \binom{n}{p n}^2 q^{-(1-p) (2L-l-1) n }.
\end{align*}

Finally, plugging in this estimate in (\ref{eqn::varexpr-B3}),
\begin{align*}
\Var~W
& \leq
\sum_{l = 1}^L L^{2L} q^{k(2L - l)} \cdot q^{-(1-p) (2L - l -1)}
    \binom{n}{np}^2
\\ &= (\E~W)^2 \cdot \sum_{l = 1}^L 4L^{2L} \cdot
        q^{((1-p) n - k) l - (1-p) n}
\\ &\leq (\E~W)^2 \cdot 4 L^{2L+1} \cdot q^{\gamma n L - (1-p) n}
.\end{align*}
Thus, for $L := \frac{1-p}{2 \gamma}$, the variance of $W$ is $o((\E~W)^2)$,
and hence we are done.

% ===============

\subsection{Proof of part~\ref{itm::reras-lin} of Theorem~\ref{thm::reras}
(random linear codes under erasures)}
    \label{apdx::subsec::reras-lin}

We first note that if a linear code contains a list of $L$ {\em linearly
independent} codewords agreeing with some $a \in {\cal A}$, then its list-size
is at least $q^{L-1}$. Indeed, if $c_1, c_2, \ldots, c_L$ are codewords
agreeing with $a$, then, in fact, so does every {\em `affine' linear
combination} of the codewords; i.e., every vector of the form $\theta_1 c_1
+ \cdots + \theta_L c_L$ where the $\theta_i$ are scalars satisfying
$\theta_1 + \theta_2 + \cdots + \theta_L = 1$. Note that the number of such
linear combinations is exactly $q^{L-1}$.

Consider a random linear code $C$ of blocklength $n$ and dimension
$k = (1-p-\gamma)n$. Recall that ${\cal A}$ is the set of strings $a$ over
$\F_q \cup \{\erasym\}$ such that $|\suppstar(a)| = (1-p)n$. For
$a \in {\cal A}$ and any linearly independent $L$-list $X$ of messages,
define $\I(a, X)$ to be the indicator random variable for the event that
$C(x)$ agrees with $a$ for all $x \in X$, and let $W := \sum_{a, X} \I(a, X)$.

For fixed $a$ and $X$, it is easy to see that $\E~\I(a, X) =
\exp_q \left( -(1-p) L n \right)$, and therefore (as in
Appendix~\ref{apdx::subsec::reras-gen}),
\[
\E~W \geq
q^{- (1-p) Ln} \cdot \binom{n}{np} q^{(1-p) n} \cdot \frac{1}{2} q^{k L}
.\]
For a pair of lists $X$ and $Y$ (each of which is linearly independent),
define $l = l(X, Y) :=  \dim(\vecspan(X)
\cap \vecspan(Y))$. It is easy to check that if $l = 0$ (i.e., $X$ and
$Y$ are linearly independent), the random variables
$\I(a, X)$ and $\I(b, Y)$ are statistically independent. Therefore, we can
bound the variance of $W$ by
\[
    \Var~W \leq \sum_{l=1}^L \sum_{l(X, Y) = l} \sum_{a,b}
        \P~[{\cal E}]
,\]
where ${\cal E}$ is the event that $\I(a, X) = 1 \text{ and } \I(b, Y) = 1$.

Fix a pair $X, Y$ such that $\dim ( \vecspan\ X \cap \, \vecspan\ Y ) = l >0$.
As in Subsection~\ref{apdx::subsec::rerro-lin}, we partition $Y$ as $Y_0
\cup Y_1$, where (a) $|Y_0| = l$ and $|Y_1| = L-l$, (b) $X$ is
linearly independent from $Y_1$, and (c) $Y_0 \subseteq \vecspan (X \cup \
Y_1)$. Moreover, pick $y_0 \in Y_0$ arbitrarily, so that $y_0 =
\sum_{u \in X \cup Y_1} \theta_u \cdot u$ for some scalars
$\{ \theta_u \}_{u \in X \cup Y_1}$. Note that $\theta_x \neq 0$ for at least
one $x \in X$.

Now, fix a pair of strings $a, b \in {\cal A}$, and let $S := \suppstar(a)$
and $T := \suppstar(b)$. We are interested in the probability of ${\cal E}$
for this choice of $a$ and $b$. (Note that for general codes, this event
implies that the strings $a$ and $b$ had to agree with each other; this is not
so for linear codes.) For any $x \in X$, conditioned on the event that
$C(x)|_S = a|_S$, the random variable $C(x)|_{T \smallsetminus S}$ is
uniformly distributed over $\F_q^{T \smallsetminus S}$.
Since $y_0 = \sum_{x \in X} \theta_x \cdot x
+ \sum_{y \in Y_1} \theta_y \cdot y$ (with $\theta_x \neq 0$ for some $x \in
X$), it follows that $C(y_0)|_{T \smallsetminus S}$ is also uniformly
distributed over $\F_q^{T \smallsetminus S}$. Hence, conditioned on the
event that $C(x)$ agrees with $a$ for all $x \in X$ and $C(y)$ agrees with
$b$ for all $y \in Y_1$, the probability that $C(y_0)$ agrees with $b$ is at
most $q^{-|T \smallsetminus S|}$. Hence,
\begin{align*}
\sum_{a,b} \P~[{\cal E}]
&\leq
\sum_{S, T} q^{-(1-p) n |X \cup Y_1|}
      q^{-|T \smallsetminus S|} \cdot q^{|S|+|T|},
\\ &=
q^{-(1-p) n (2L - l)} q^{2(1-p)n} \cdot \sum_{S, T}
      q^{-|T \smallsetminus S|},
\\ &\leq
q^{-(1-p) n (2L - l)} q^{2(1-p)n} \cdot \sum_{S, T}
      2^{-|T \smallsetminus S|},
\\ &=
q^{-(1-p) n (2L-l)} q^{2(1-p) n} \binom{n}{np}^2
    \E_{S,T}~\left[ 2^{-|T \smallsetminus S|} \right].
\end{align*}
Here, the expectation is over $S,T \subseteq [n]$ of size $(1-p) n$, chosen
independently and uniformly at random. By Lemma~\ref{lem::expexpHG}
below, this quantity can be bounded by
\begin{align*}
q^{-(1-p) n (2L-l)} q^{2(1-p) n} \binom{n}{np}^2 \cdot
    \exp_2 \left( -\frac{1}{8} p(1-p)n + o(n) \right)
.\end{align*}
Plugging this in our upper bound for the variance, we have
\begin{align*}
\Var~W
&\leq
\sum_{l=1}^L q^{2Ll} q^{k(2L-l)} \cdot q^{-(1-p) n (2L-l)} q^{2(1-p) n}
    \binom{n}{np}^2 \cdot
    2^{-\frac{1}{8} p(1-p)n + o(n)}
\\ &\leq
4 (\E~W)^2 \sum_{l=1}^L q^{2Ll} q^{((1-p) n - k) l} \cdot
    2^{-\frac{1}{8} p(1-p)n + o(n) }
\\ &\leq
4 L q^{2L^2} q^{\gamma n L}
    2^{-\frac{1}{8} p(1-p)n + o(n)} \cdot (\E~W)^2
\end{align*}
Thus, for \[
L := \frac{p(1-p)} {16 \gamma \log q} ,\]
this ratio is $o(1)$. Thus, the code contains a bad list of $L$ {\em
linearly independent} messages w.h.p.; this implies that its list-size is
at least $q^{L-1}$.

\begin{lemma}
\label{lem::expexpHG}
If $S, T$ are independently and uniformly random subsets of $[n]$ of size
$(1-p) n$, then \[ \E_{S,T} \left[ 2^{-|T \smallsetminus S|} \right] \leq
\exp_2 \left( -\frac{p(1-p) n}{8} + o(n) \right) .\]
\end{lemma}

\begin{proof}
We prove this by thresholding on $|T \smallsetminus S|$. It can be easily
checked that the random variable
$|T \smallsetminus S|$ has the hypergeometric distribution with parameters
$(n, pn, (1-p)n)$, and hence its mean is $p(1-p) n$. Hence, since
hypergeometric random variables are concentrated around their mean, we expect
that $|T \smallsetminus S| \geq \frac{1}{8} p(1-p)$, except with an
exponentially small probability.

We now justify the above intuition by explicit calculations. For any $t$, the
probability that $|T \smallsetminus S| =
t$ is equal to
\[
 f(n, (1-p)n, pn, t) := \frac{\binom{(1-p)n}{t} \binom{pn}{pn-t}}
{\binom{n}{pn}} = \frac{\binom{(1-p)n}{t} \binom{pn}{t}}
{\binom{n}{pn}}.
\]
For $t \leq \frac{1}{8} p(1-p) n$, this can be upper bounded by
$2^{\eps n + o(n)}$, where
\[
\eps := (1-p) \h\left( \frac{p}{8} \right) + p \h\left( \frac{1-p}{8} \right)
        - \h(p).
\]
We are interested in upper bounding the exponent $\eps$. We will assume
that $p \leq 1/2$; the argument in the $p \geq 1/2$ case is symmetric (by
replacing $p$ by $1-p$). By concavity of $\h(\cdot)$,
\[
\eps
\leq \h\left( (1-p) \cdot \frac{p}{8} + p \cdot \frac{1-p}{8} \right) - \h(p)
\leq \h(p/4) - \h(p).
\]
By Fact~\ref{fact::htaylor},
\[
\eps \leq
\left[ \frac{p}{4} \log \left( \frac{4}{p} \right) -
    p \log \left( \frac{1}{p} \right) \right]
+ (\log e) \left[ \frac{p}{4} - (p - p^2) \right]
.\]
For $0 < p \leq 1/2$, the first term is negative, and hence
$\eps \leq (\log e) \left( p^2 - \frac{3p}{4} \right) \leq
-\frac{1}{4} p \log e \leq -\frac{p(1-p)}{4}$.

Thus, summing over all $t \leq \frac{1}{8} p(1-p)n$, the event
$|T \smallsetminus S| \leq \frac{1}{8} p(1-p)n$ occurs with
probability at most $\exp_2 \left( - \frac{p(1-p)n}{4} + o(n) \right)$. Hence,
the desired expectation is bounded as
\begin{align*}
\E \left[ 2^{-|T \smallsetminus S|} \right]
&\leq
\P \left[ |T \smallsetminus S| \leq \frac{1}{8} p(1-p)n \right] \cdot 1
+ \exp_2 \left( -\frac{1}{8} p(1-p)n \right),
\end{align*}
establishing the claim.
\end{proof}

\end{document}